\documentclass[10pt,letterpaper]{article}
\pdfoutput=1

\usepackage{jcappub}
\usepackage{amsmath,amssymb,color}
\usepackage{enumerate,xstring}
\usepackage{jcappub}
\usepackage{amsmath}
\usepackage{amsfonts}
\usepackage{amssymb}
\usepackage{graphicx}
\usepackage{epsfig}
\usepackage{color}
\usepackage{multirow}
\usepackage{graphicx}
\usepackage{bigints}
\usepackage{todonotes,bm,etoolbox}
\usepackage{calligra}
\usepackage{hyperref}
\usepackage{tabularx}
\usepackage{booktabs}

\usepackage{tikz}

\def\ba#1\ea{\begin{align}#1\end{align}}
\def\bea{\begin{eqnarray}}
\def\eea{\end{eqnarray}}
\def\be{\begin{equation}}
\def\ee{\end{equation}}

\def\({\left(}
\def\){\right)}
\def\[{\left[}
\def\]{\right]}

\def\<{\left\langle}
\def\>{\right\rangle}

\def\comment#1{}

\def\eps{\epsilon}

\renewcommand{\v}[1]{\bm{#1}}

\def\vx{\v{x}}
\def\vk{\v{k}}
\def\vq{{\v{q}}}

\def\vv{\v{v}}

\newcommand{\perm}[1]{ \expandafter\ifstrempty\expandafter{#1} {\mbox{perm.}} {\mbox{$#1$ perm.}} }

\def\O{\mathcal{O}}

\newcommand{\fnl}{f_\textnormal{\textsc{nl}}}

\definecolor{RedWine}{rgb}{0.743,0,0}
\definecolor{RoyalBlue}{rgb}{0.25,.41,.88}
\definecolor{ForestGreen}{rgb}{.13,.54,.13}
\definecolor{Goldenrod}{rgb}{.85,.65,.13}

\newcommand{\bq}{\begin{eqnarray}}
\newcommand{\eq}{\end{eqnarray}}

\title{\huge Galaxy bias and primordial non-Gaussianity: \Large insights from galaxy formation simulations with IllustrisTNG}

\author[a]{Alexandre Barreira,}
\author[a]{Giovanni Cabass,}
\author[a]{Fabian Schmidt,}
\author[b]{Annalisa~Pillepich}
\author[a]{and Dylan Nelson}

\affiliation[a]{Max-Planck-Institut f\"ur Astrophysik, Karl-Schwarzschild-Stra\ss e~1, 85748 Garching, Germany}
\affiliation[b]{Max-Planck-Institut f\"ur Astronomie, K\"onigstuhl 17, 69117 Heidelberg, Germany}

\emailAdd{barreira@mpa-garching.mpg.de}
\emailAdd{gcabass@mpa-garching.mpg.de}
\emailAdd{dnelson@mpa-garching.mpg.de}
\emailAdd{fabians@mpa-garching.mpg.de}
\emailAdd{pillepich@mpia-hd.mpg.de}

\date{\today}

\abstract{We study the impact that large-scale perturbations of (i) the matter density and (ii) the primordial gravitational potential with local primordial non-Gaussianity (PNG) have on galaxy formation using the IllustrisTNG model. We focus on the linear galaxy bias $b_1$ and the coefficient $b_\phi$ of the scale-dependent bias induced by PNG, which describe the response of galaxy number counts to these two types of perturbations, respectively. We perform our study using separate universe simulations, in which the effect of the perturbations is mimicked by changes to the cosmological parameters: modified cosmic matter density for $b_1$ and modified amplitude $\mathcal{A}_s$ of the primordial scalar power spectrum for $b_\phi$. We find that the widely used universality relation $b_\phi = 2\delta_c(b_1 - 1)$ is a poor description of the bias of haloes and galaxies selected by stellar mass $M_*$, which is instead described better by $b_\phi(M_*) = 2\delta_c(b_1(M_*) - p)$ with $p \in [0.4, 0.7]$. This is explained by the different impact that matter overdensities and local PNG have on the median stellar-to-halo-mass relation. A simple model of this impact allows us to describe the stellar mass dependence of $b_1$ and $b_\phi$ fairly well. Our results also show a nontrivial relation between $b_1$ and $b_\phi$ for galaxies selected by color and black hole mass accretion rate. Our results provide refined priors on $b_\phi$ for local PNG constraints and forecasts using galaxy clustering. Given that the widely used universality relation underpredicts $b_\phi(M_*)$, existing analyses may underestimate the true constraining power on local PNG.}

\begin{document}

\maketitle

\section{Introduction}
\label{sec:introduction}

The study of {\it galaxy bias}, i.e., the connection between the observed galaxy distribution and the properties of the underlying matter distribution, is a long-standing active field of research in cosmology and astrophysics (see e.g.~Ref.~\cite{biasreview} for a comprehensive review). In full generality, the number density of galaxies $n_g(\vx, z)$ in their rest-frame at position $\vx$ and redshift $z$ in the Universe can be written as
\bq\label{eq:biasexp}
n_g(\vx, z) = \bar{n}_g(z) \left[1 + \sum_{\O}b_\O(z)\O(\vx, z) + \eps(\vx)\right],
\eq
where $\bar{n}_g(z)$ is the number of galaxies averaged over all positions at redshift $z$ and the sum runs over all types of perturbations of the mass and energy distribution of the Universe $\O(\vx, z)$ that influence galaxy formation. The coefficients $b_\O(z)$ are called the {\it galaxy bias parameters}, they depend on properties like galaxy mass, color, star formation rate, etc., and formally describe how the number of galaxies around position $\vx$ changes when the amplitude of $\O$ changes at $\vx$. If one restricts to perturbations $\O$ that are sufficiently large-scale and amenable to analytic perturbation theory description \cite{Bernardeau/etal:2002}, then all of the complications of galaxy formation are absorbed by the finite set of bias parameters, which can be determined from the observational data. Galaxy formation still depends on smaller-scale fluctuations of the matter/energy distribution that, at first order, do not correlate with the large-scale perturbations $\O$; in Eq.~(\ref{eq:biasexp}), $\eps(\vx)$ is one of these so-called {\it stochastic} (or shot-noise) contributions.

Within the formalism of Eq.~(\ref{eq:biasexp}), the first step is to enumerate all of the relevant perturbations $\O$. For example, total matter density perturbations $\delta_m(\vx, z)$ contribute to linear order as $b_1(z)\delta_m(\vx, z)$ \cite{fry/gaztanaga:1983, lazeyras/etal}, where $b_1(z)$ is the popular linear local-in-matter-density (LIMD) galaxy bias. Additional types of perturbations include large-scale tidal fields, $\O= [K_{ij}(\vx, z)]^2$ \cite{mcdonald/roy:2009, chan/scoccimarro/sheth:2012, baldauf/etal:2012, saito/etal:14, 2018JCAP...09..008L, 2018JCAP...07..029A}; higher-derivative bias terms, $\O = \nabla^2\delta_m(\vx, z)$ \cite{2019JCAP...11..041L}; baryon-cold dark matter (CDM) relative density $\O = \delta_{bc}(\vx, z)$ \cite{barkana/loeb:11, 2016PhRvD..94f3508S, 2016ApJ...830...68A} and velocity $\O = \{\nabla \vv_{bc}, \vv_{bc}^2(\vx, z)\}$ \cite{tseliakhovich/hirata:2010, blazek/etal:15, 2016PhRvD..94f3508S} perturbations generated by photon-baryon interactions before the epoch of decoupling; modulations of the gravitational potential $\O  = \phi(\vx)$ in primordial non-Gaussianity (PNG) contexts \cite{dalal/etal:2008, 2010CQGra..27l4011D} and primordial compensated baryon-CDM isocurvature perturbations \cite{2020JCAP...02..005B, 2019PhRvD.100j3528H, 2020arXiv200212931B}. Predicting the values of the corresponding bias parameters themselves is very challenging. Yet it is important for at least two reasons. First, even though analyses of galaxy clustering data (e.g.~Refs.~\cite{2017MNRAS.470.2617A, 2017MNRAS.464.1640S, 2019arXiv190905277I, 2019arXiv190905271D} for recent analyses of the BOSS DR12 galaxy data) effectively marginalize over the $b_\O(z)$ to obtain constraints on the cosmological parameters, a good theoretical knowledge of galaxy bias is still important to inform the priors over which to sample/marginalize the $b_\O(z)$, guide modeling of their time-evolution (important for tomographic analyses), and help establish hierarchies and relations between them to reduce the number of free parameters. The second reason has to do with the fact that there is a great deal that could be learned about the astrophysical processes that govern galaxy formation (like gas accretion/cooling, star formation/feedback, black hole growth/feedback) if observational determinations of galaxy bias, which encode the dependence of galaxy formation on the long-wavelength environment, could be compared with predictions from different models of the formation and evolution of galaxies.

Theoretical predictions of galaxy bias are however extremely challenging to obtain because of the complex interaction of the many physical processes that are at play. For this reason, most theoretical studies to date have focused on {\it halo bias} in the context of collisionless (i.e., gravity-only) structure formation dynamics. The most widely studied are the LIMD bias parameters $\propto b_n(z)\delta_m(\vx, z)^n$, for which accurate and popular fitting formulae from gravity-only simulations exist \cite{sheth/tormen:1999, 2010ApJ...724..878T, lazeyras/etal}, but tidal and other higher-order bias parameters have been extensively studied for haloes as well \cite{chan/scoccimarro/sheth:2012, baldauf/etal:2012, saito/etal:14, 2018JCAP...09..008L, 2018JCAP...07..029A, 2019JCAP...11..041L}. {\it Galaxy bias} is far less well studied because of the numerical challenges of carrying out sufficiently high-resolution hydrodynamical simulations in sufficiently large volumes. Fortunately, in recent years, these challenges have started to be overcome with projects like {\sc Illustris} (box size $L_{\rm box} \approx  100\ {\rm Mpc}$) \cite{2014MNRAS.444.1518V}, {\sc EAGLE} ($L_{\rm box} = 100\ {\rm Mpc}$) \cite{2015MNRAS.446..521S, 2017arXiv170609899T}, {\sc Magneticum} ($L_{\rm box} = 500\ {\rm Mpc}$; note this project spans a range of volumes and resolutions) \cite{2014MNRAS.442.2304H}, {\sc BAHAMAS} ($L_{\rm box} \approx 570\ {\rm Mpc}$) \cite{2017MNRAS.465.2936M}, {\sc Horizon-AGN} ($L_{\rm box} \approx 142\ {\rm Mpc}$) \cite{2014MNRAS.444.1453D} and {\sc IllustrisTNG} ($L_{\rm box} \approx 50, 100, 300\ {\rm Mpc}$) \cite{Pillepich:2017jle, 2017MNRAS.465.3291W, Nelson:2018uso}. In these simulations, structure formation takes place in cosmological volumes under the action of gravity and hydrodynamics, as well as physical processes such as star formation, and stellar and black hole feedback that are implemented as coarse-grained effective models with parameters that are calibrated to reproduce a handful of observations (e.g.~the stellar mass function at low redshift, the cosmic star formation rate history or cluster gas fractions). These simulations therefore make it possible to begin to systematically study the actual {\it galaxy} bias (not just halo bias), and as a function of properties that are more directly measurable in observations such as stellar mass, luminosity or star formation rate (and not just total halo mass). The advent of these simulations also opens the door to study the bias parameters associated with perturbations $\O$ that explicitly distinguish between the distribution of baryons and CDM, such as the $\delta_{bc}$, $\vv_{bc}$ and the compensated isocurvature perturbations mentioned above \cite{2020JCAP...02..005B}.

In this paper, we use simulations of the IllustrisTNG galaxy formation model to study two particularly important galaxy bias parameters: the linear LIMD parameter $b_1$ and the bias parameter $b_\phi$ associated with primordial non-Gaussianity of the local type. The latter is parametrized in terms of the primordial gravitational (Bardeen) potential during matter domination $\phi(\vx)$ and the parameter $\fnl$ as \cite{2001PhRvD..63f3002K}
\bq\label{eq:fnl}
\phi(\vx) = \phi_{\rm G}(\vx) + \fnl\left[\phi_{\rm G}(\vx)^2 - \left<\phi_{\rm G}(\vx)^2\right>\right],
\eq
where $\phi_{\rm G}$ is a Gaussian distributed random field and $\left<\cdots\right>$ denotes ensemble average. The simplest single-field models of inflation predict vanishing $\fnl$ \cite{2011JCAP...11..038C, Tanaka:2011aj, baldauf/etal:2011, conformalfermi, CFCpaper2, 2015JCAP...10..024D}, and hence, any detection of a non-zero $\fnl$ would carry immediate and very far-reaching consequences for our knowledge of the early Universe and the mechanism that generated the seeds of structure formation. The current best constraints on local PNG come from the analysis of the cosmic microwave background (CMB) by the Planck satellite, which set  $\fnl = -0.9 \pm 5.1\ (1\sigma)$ \cite{2019arXiv190505697P}. The galaxy distribution can also be a very powerful probe of local PNG via the so-called {\it scale-dependent bias} feature in the galaxy power spectrum \cite{dalal/etal:2008}. Concretely, in such studies, the relevant galaxy bias expansion is (see e.g.~Refs.~\cite{mcdonald:2008, PBSpaper, angulo/etal:2015, 2015JCAP...11..024A, assassi/baumann/schmidt} for discussions about the contribution from PNG to the galaxy bias expansion)
\bq\label{eq:biasexp_2}
\delta_g(\vx, z)  = b_1(z)\delta_m(\vx, z) + b_\phi(z)\fnl\phi(\vx) + \eps(\vx),
\eq
where $\delta_g(\vx, z) = n_g(\vx, z)/\bar{n}_g(z) - 1$. The corresponding galaxy power spectrum (Fourier transform of the two-point correlation function) is defined as $(2\pi)^3P_{gg}(k,z)\delta_D(\vk+\vk') = \langle\delta(\vk, z)\delta(\vk',z)\rangle$ with (omitting the redshift dependence of the bias parameters)
\bq\label{eq:Pgg}
P_{gg}(k,z) &=& b_1^2P_{mm}(k,z) + 2b_1b_\phi \fnl P_{m\phi}(k,z) + b_\phi^2 \fnl^2 P_{\phi\phi}(k) + P_{\eps\eps}(k) \nonumber \\
&=& \left[b_1^2 + \frac{2b_1b_\phi \fnl}{\mathcal{M}(k,z)} + \frac{b_\phi^2 \fnl^2}{\mathcal{M}(k,z)^2}\right] P_{mm}(k,z) + P_{\eps\eps},
\eq
where $P_{ab}$ denotes the cross-power spectrum of the fields $a, b$ ($m \equiv \delta_m$) and $P_{\eps\eps}$ is the $k$-independent power spectrum of the noise. In the second line of Eq.~(\ref{eq:Pgg}) we have used the relation $\delta_m(\vk, z) = \mathcal{M}(k,z)\phi(\vk)$, where
\bq\label{eq:M}
\mathcal{M}(k,z) = \frac{2}{3}\frac{k^2T_m(k,z)}{\Omega_{m0}H_0^2},
\eq
and $T_m$ is the matter transfer function, $\Omega_{m0}$ is the fractional matter density parameter today and $H_0$ is the Hubble expansion rate today. On scales $k \lesssim 0.01\ h/{\rm Mpc}$, the transfer function is scale-independent and thus $\fnl$ induces scale-dependent corrections $\propto b_1b_\phi\fnl/k^2$ and $\propto b_\phi^2\fnl^2/k^4$ relative to $P_{mm}$ that can be used to put bounds on $\fnl$. The amplitude of the effect is however completely degenerate with the bias parameter $b_\phi$, and consequently, searches for PNG using the galaxy distribution therefore depend critically on our knowledge of $b_\phi$.

Assuming universality of the halo mass function, it can be shown that \cite{slosar/etal:2008, 2008ApJ...677L..77M, afshordi/tolley:2008, 2010A&A...514A..46V, matsubara:2012, ferraro/etal:2012, PBSpaper, scoccimarro/etal:2012, 2017MNRAS.468.3277B}
\bq\label{eq:bphiuniv}
b_\phi(z) = 2\delta_c\left(b_1(z) - 1\right),
\eq
where $\delta_c = 1.686$ is the (linearly extrapolated to $z=0$) threshold overdensity for spherical collapse. When compared to estimates obtained for haloes in gravity-only $N$-body simulations \cite{grossi/etal:2009, desjacques/seljak/iliev:2009, 2010MNRAS.402..191P, 2010JCAP...07..013R, 2011PhRvD..84h3509H, scoccimarro/etal:2012, 2012JCAP...03..002W, baldauf/etal:2015, 2017MNRAS.468.3277B}, this {\it universality relation} is found to provide a decent first order approximation for $b_1 \lesssim 1.5$, but to overpredict simulation results for $b_1 \gtrsim 1.5$; the exact level of the overprediction depends on halo definition (see e.g.~Ref.~\cite{2017MNRAS.468.3277B} for a recent discussion). This relation (or slight variations of it) is commonly adopted in existing searches for local PNG using the galaxy power spectrum \cite{slosar/etal:2008, 2011JCAP...08..033X, 2013MNRAS.428.1116R, 2014PhRvD..89b3511G, 2014PhRvL.113v1301L, 2014MNRAS.441L..16G, 2015JCAP...05..040H, 2019JCAP...09..010C}, as well as in forecast studies of the expected constraining power of future galaxy data \cite{2008ApJ...684L...1C, 2012MNRAS.422.2854G, 2014arXiv1412.4872D, 2014arXiv1412.4671A, 2015JCAP...01..042R, 2015PhRvD..92f3525A, 2015MNRAS.448.1035C, 2017PhRvD..95l3513D, 2017PDU....15...35R}. Uncertainties on our theoretical understanding of $b_\phi$ translate therefore in equally large uncertainties on the resulting PNG constraints obtained with the galaxy power spectrum. Our goal in this paper is to go beyond gravity-only predictions and contribute to an improved understanding of $b_\phi$ and its relation to $b_1$ for galaxies as a function of more directly observable quantities like stellar mass or color.

Concretely, in this paper we present separate universe simulations of the IllustrisTNG model to study both $b_1$ and $b_\phi$ for simulated galaxies in a cosmological context. The separate universe technique is an efficient numerical method to predict galaxy bias that takes advantage of the equivalence between the response of galaxy formation to long-wavelength perturbations and the response of galaxy formation to changes in the background cosmology. Our results will show that Eq.~(\ref{eq:bphiuniv}) is not a good description of $b_\phi$ for stellar-mass selected objects: it underpredicts the measured $b_\phi$ for all of the $b_1$ (and corresponding mass scales) probed and we find that $2\delta_c(b_1 - p)$ with $p \in [0.4, 0.7]$ is a more appropriate description. This breakdown of the universality relation can be traced back to the different impact that matter ($\delta_m$) and potential ($\fnl\phi$) perturbations have on the stellar-to-halo-mass relation (SHMR). We will also look briefly into the dependence of the bias parameters $b_1$ and $b_\phi$ on galaxy color and black hole mass accretion rate. 

To the best of our knowledge, our results on $b_\phi$ constitute the first predictions for fully self-consistently simulated galaxies and can be used to improve the theoretical priors currently employed in constraint/forecast analysis of local PNG.\footnote{A few past numerical studies using hydrodynamical simulations of cosmologies with local PNG exist \cite{2011MNRAS.415.3021M, 2011CQGra..28v5015M, 2012MNRAS.421.1113M, 2013arXiv1307.5051Z}, but their numerical setups do not permit to measure the galaxy bias $b_\phi$.}

The outline of this paper is as follows. We describe our numerical simulations in Sec.~\ref{sec:sepuni} and, in Sec.~\ref{sec:biasresults}, we present our results for $b_1$ and $b_\phi$ as a function of total mass $M_{\rm h}$, stellar mass $M_*$, galaxy color ($g-r$) and black hole mass accretion rate $\dot{M}_{\rm BH}$. The impact of matter overdensities and local PNG on the SHMR relation is discussed in Sec.~\ref{sec:shmrimpact}, where we also describe a simple model of the stellar mass dependence of $b_1$ and $b_\phi$. We summarize and conclude in Sec.~\ref{sec:summary}.

\section{Separate Universe simulations}
\label{sec:sepuni}

\begin{figure}[t!]
	\centering
	\includegraphics[width=\textwidth]{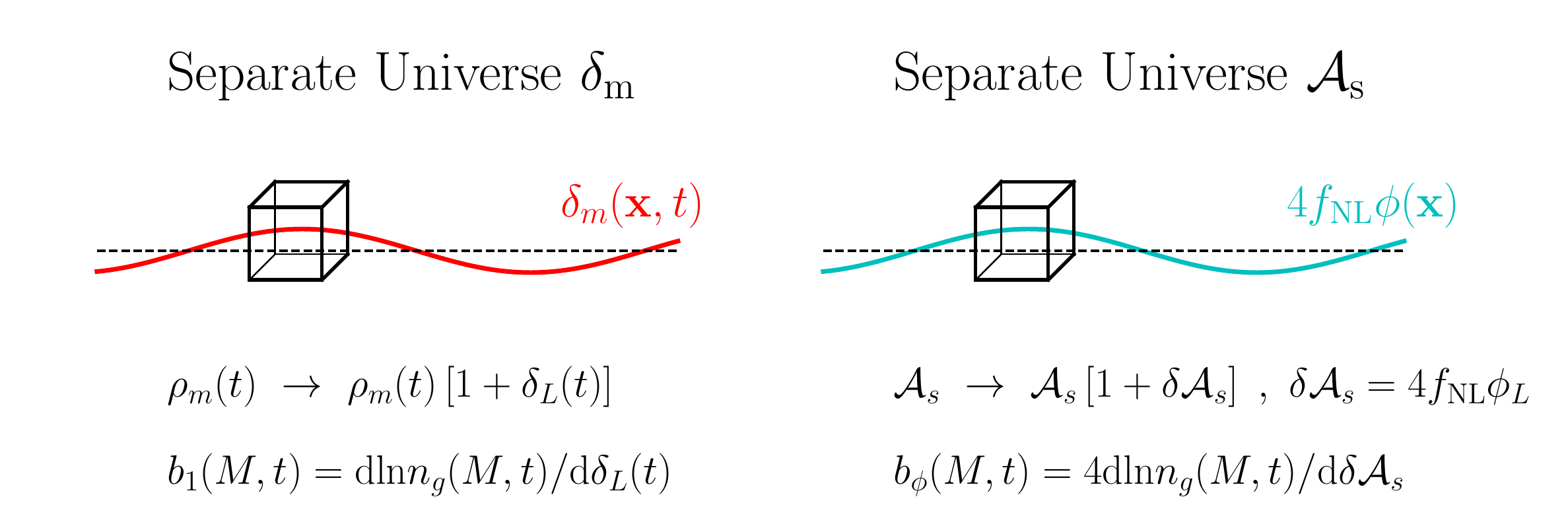}
	\caption{Summary sketch of the separate universe approach to predict the galaxy bias parameters $b_1$ and $b_\phi$ in Eq.~(\ref{eq:biasexp_2}). Local structure formation in a fiducial cosmology inside long-wavelength matter perturbations $\delta_m$ is equivalent to global stucture formation in a separate cosmology with modified background matter density ${\rho}_m$. Similarly, local structure formation in a fiducial cosmology inside long-wavelength primordial gravitational potential perturbations with PNG of the local type is equivalent to global structure formation in a cosmology with modified amplitude of the primordial scalar perturbation power spectrum $\mathcal{A}_s$. The bias parameters $b_1$,$b_\phi$ are evaluated as the {\it response} of galaxy number counts to the amplitudes of the corresponding perturbations $\delta_L$, $\delta \mathcal{A}_s = 4\fnl \phi_L$, respectively.} 
\label{fig:idea}
\end{figure}
In this section we describe the simulations that we perform to measure the linear LIMD galaxy bias parameter $b_1$ and the bias parameter $b_\phi$ associated with PNG of the local type that enter the galaxy bias expansion of Eq.~(\ref{eq:biasexp_2}). The parameters $b_1$ and $b_\phi$ describe, respectively, the {\it response} of galaxy number counts to long-wavelength $\delta_m(\vx, z)$ and $\phi(\vx)$ perturbations with local PNG. Under the assumption that the physics of galaxy formation acts on sufficiently small scales relative to the size of the perturbations, the latter effectively act as a modified background to the galaxies forming on small scales (this is called the {\it peak-background split} argument \cite{kaiser:1984, bardeen/etal:1986}). One can then invoke the separate universe argument, which states that:

\vspace{0.2cm}

{\it Local structure formation inside long-wavelength perturbations in a fiducial cosmology is equivalent to global structure formation at cosmic mean in an appropriately modified cosmology.} 

\vspace{0.2cm}

For the case of $b_1$, the change in cosmology corresponds to changing the background matter density, whereas for $b_\phi$, the modified cosmology has a different amplitude of the primordial scalar power spectrum, $\mathcal{A}_s$. Next, we describe the numerical details of the simulations of our fiducial cosmology, and then discuss which aspects of it differ for the case of the modified ones. Figure \ref{fig:idea} and Table \ref{table:params} provide a summary of the separate universe approach and the cosmologies we consider in this paper.

\subsection{Simulations and identification of structure}
\label{sec:fiducial}

All of the simulations presented in this work are carried out with the moving-mesh hydrodynamic + gravity $N$-body code {\sc AREPO} \citep{2010MNRAS.401..791S, 2016MNRAS.455.1134P} together with the IllustrisTNG model \citep{2017MNRAS.465.3291W, Pillepich:2017jle,Nelson:2018uso}. The latter is an effective physical model for galaxy formation and evolution in a cosmological setup that includes prescriptions for star formation, stellar feedback, chemical enrichment, galactic winds, gas cooling, gas (re)ionization, and black hole seeding and growth with different feedback modes at different accretion rates. The IllustrisTNG model is an improved version of the Illustris model \citep{2014MNRAS.445..175G, 2014MNRAS.444.1518V} and it broadly reproduces a number of observations including the low redshift galaxy stellar mass function, the star formation rate history, the stellar-to-halo-mass relation and the gas fractions in galaxies and galaxy groups. The interested reader is referred to Refs.~\cite{2018MNRAS.480.5113M, Pillepich:2017fcc, 2018MNRAS.477.1206N, 2018MNRAS.475..676S, Nelson:2017cxy} for the first results with IllustrisTNG.

Our simulations start at redshift $z_i = 127$ with initial conditions generated by the {\sc N-GenIC} code \citep{2015ascl.soft02003S} using the Zel'dovich approximation. The linear matter power spectrum given to {\sc N-GenIC} is obtained by rescaling the result of the {\sc CAMB} code \citep{camb} at $z=0$ back to $z_i$ assuming no cosmic radiation density in the growth factor (consistently with the subsequent forward evolution by {\sc Arepo}). We perform simulations at two numerical resolutions. One, which we label as TNG300-2, comprises $N_p = 1250^3$ dark matter mass elements and $N_p = 1250^3$ initial gas elements in a cubic box with size $L_{\rm box} = 205\ {\rm Mpc}/h$. The other one is higher resolution, labeled as TNG100-1.5, and it contains the same number of mass elements but in a smaller box size $L_{\rm box} = 75\ {\rm Mpc}/h$. For each resolution, the initial conditions of all of the cosmologies were generated with the same random white-noise seed in {\sc N-GenIC}; further, in addition to the simulations with the full IllustrisTNG model (dubbed Hydro below), we have run also gravity-only counterparts (dubbed Gravity below).\footnote{{The Hydro simulations at TNG100-1.5 and TNG300-2 resolutions were completed in approximately 1.5M CPU hours on 5120 computer cores and 0.9M CPU hours on 2560 computer cores, respectively. Note however that these figures can vary by factors of a few depending on the exact specifications of the model, code, number of cores and scaling configuration. The interested reader can consult Table A.1 of Ref.~\cite{Nelson:2017cxy} for the numbers at other mass resolution/volumes.}} To complement and extend our analysis of $b_1$ and $b_\phi$ as a function of total halo mass to higher mass values, we have run additional gravity-only simulations with the same number of mass elements but with a bigger box size, $L = 560\ {\rm Mpc}/h \approx 800\ {\rm Mpc}$ (labeled as $L\approx 800\ {\rm Mpc}$; these have no Hydro counterpart). 

We will show measurements of the bias for both haloes and subhaloes. The haloes correspond to structures identified by a Friends-of-Friends (FoF) algorithm run on the dark matter mass elements with linking length $b = 0.2$ times the mean interparticle distance in the simulations. The subhaloes correspond in turn to gravitationally bound substructures found by the {\sc SUBFIND} algorithm \cite{2001MNRAS.328..726S} inside each halo. The subhalo sample includes the main {\it central} subhaloes that reside at the bottom of the potential well of their parent halo, as well as the remaining {\it satellite} subhaloes that orbit around the central subhalo. In all of our results, when we quote the value of a given quantity for a given object (halo or subhalo), we always consider that quantity using all of the mass elements that belong to the object; for example, the stellar mass of a halo/subhalo is the summed mass of all star particles that belong to that halo/subhalo (for the case of a halo this includes the stars inside all its subhaloes).  In the case of the Hydro simulations, we consider only haloes and subhaloes that contain stars, i.e., ($M_* > 0$); in our nomenclature, we refer to these subhaloes as galaxies and we make no distinction between central and satellite galaxies. 

We emphasize that, when we adjust the cosmological parameters from the fiducial to the separate universe cosmologies, we keep the parameters of the IllustrisTNG model fixed. The overall philosophy in developing galaxy formation simulations is to select a small number of key galaxy observations and calibrate the parameters of the models to reproduce them. From this point on, the results of the models for quantities or epochs that were not used in the calibration (like the galaxy bias parameters) count as predictions and can potentially be used to test the models. Our numerical results should thus be interpreted as predictions for the response of galaxy formation to the presence of long-wavelength density perturbations (for $b_1$) and local PNG (for $b_\phi$), at fixed galaxy formation physics prescription (that of IllustrisTNG). In fact, it is conceivable to expect robust comparisons between galaxy bias estimated from data and theoretical predictions like ours here to become possible in the future; this could add another line of testing of galaxy formation models in their cosmological predictions and contribute to advances in the study of galaxy formation that way (see e.g.~Fig.~11 of Ref.~\cite{2018MNRAS.475..676S} for a demonstration of how IllustrisTNG galaxies recover well the observed relative clustering amplitude (i.e., bias) between red and blue galaxies at low redshift). 
  
\subsection{The $\delta_m$ Separate Universe}
\label{sec:sepunidm}

\begin{table}
\centering
\begin{tabular}{@{}lccccccccccc}
\hline\hline
\rule{0pt}{1\normalbaselineskip}
Name &\ \ $\Omega_{m0}$ & \ \ $\Omega_{b0}$ & \ \ $\Omega_{c0}$ & \ \ $\Omega_{\Lambda0}$ & \ \ $h$ & \ \ $n_s$ & \ \ $\mathcal{A}_s$
\\
\hline
\rule{0pt}{1\normalbaselineskip}
${\rm Fiducial}$ &\ \ $0.3089$ & \ \ $0.0486$ & \ \ $0.2603$ & \ \ $0.6911$ & \ \ $0.6774$ & \ \ $0.967$ & \ \ $2.068 \times 10^{-9}$ 
\\
\\
\,\,${\rm High}\ \delta_m$ &\ \ $0.3194$ & \ \ $0.0502$ & \ \ $0.2692$ & \ \ $0.7146$ & \ \ $0.6662$ & \ \ {\footnotesize ${\rm Fiducial}$} & \ \ {\footnotesize ${\rm Fiducial}$}
\\
\,\,${\rm Low}\ \delta_m$ &\ \ $0.2991$ & \ \ $0.0471$ & \ \ $0.2520$ & \ \ $0.6691$ & \ \ $0.6884$ & \ \ {\footnotesize ${\rm Fiducial}$} & \ \ {\footnotesize ${\rm Fiducial}$}
\\
\\
\,\,${\rm High}\ \mathcal{A}_s$ &\ \ {\footnotesize ${\rm Fiducial}$} & \ \ {\footnotesize ${\rm Fiducial}$} & \ \ {\footnotesize ${\rm Fiducial}$} & \ \ {\footnotesize ${\rm Fiducial}$} & \ \ {\footnotesize ${\rm Fiducial}$} & \ \ {\footnotesize ${\rm Fiducial}$} & \ \ $2.171 \times 10^{-9}$
\\
\,\,${\rm Low}\ \mathcal{A}_s$ &\ \ {\footnotesize ${\rm Fiducial}$} & \ \ {\footnotesize ${\rm Fiducial}$} & \ \ {\footnotesize ${\rm Fiducial}$} & \ \ {\footnotesize ${\rm Fiducial}$} & \ \ {\footnotesize ${\rm Fiducial}$} & \ \ {\footnotesize ${\rm Fiducial}$} & \ \ $1.965 \times 10^{-9}$
\\
\hline
\hline
\end{tabular}
\caption{Cosmological parameter values used in the simulations of this paper. The ${\rm High}\ \delta_m$ and ${\rm Low}\ \delta_m$ cosmologies are used to measure $b_1$ and mimic the impact of long-wavelength total matter perturbations with present-day amplitudes $\delta_{L0} = +0.05$ and $\delta_{L0} = -0.05$, respectively (cf.~Sec.~\ref{sec:sepunidm}). The ${\rm High}\ \mathcal{A}_s$ and ${\rm Low}\ \mathcal{A}_s$ cosmologies are used to measure $b_\phi$ and mimic the impact of long-wavelength perturbations of the primordial gravitational potential with amplitude $\phi_L = +0.05/(4\fnl)$ and $\phi_L = -0.05/(4\fnl)$, respectively (cf.~Sec.~\ref{sec:sepuniAs}). The IllustrisTNG model was used to simulate galaxy formation in these cosmologies at two resolutions: TNG300-2 with $L_{\rm box} = 205\ {\rm Mpc/h}$ and TNG100-1.5 with $L_{\rm box} = 75\ {\rm Mpc/h}$, both with the same number of mass elements $N_p = 2\times 1250^3$; note the box sizes are rescaled slightly for the ${\rm High}\ \delta_m$ and ${\rm Low}\ \delta_m$ cosmologies (cf.~Sec.~\ref{sec:sepunidm}). For each resolution and cosmology, we have run simulations with the full IllustrisTNG model (dubbed Hydro), as well as gravity-only counterparts (dubbed Gravity). We have also a Gravity set of simulations with $N_p = 1250^3$ and $L = 560\ {\rm Mpc}/h$ (labeled as $L \approx 800\ {\rm Mpc}$).}
\label{table:params}
\end{table}

Consider galaxies forming inside a long-wavelength total matter density perturbation $\delta_m(\vx, t)$. If the size of this perturbation is sufficiently large, then the galaxies that form on smaller scales effectively regard it as a spatially constant modification of the total mass that is available locally, i.e.~the galaxies form as if they were at cosmic mean in a separate universe with a different cosmic matter density (cf.~left-hand side of Fig.~\ref{fig:idea}). Concretely, if $\delta_L(t)$ denotes the amplitude of a matter perturbation in a fiducial cosmology with background matter density $\rho_{m}(t)$, the corresponding separate universe cosmology is one with background matter density (tilded quantities correspond to the separate universe cosmology)
\bq\label{eq:rhotilde}
\tilde{\rho}_m(t) = \rho_m(t)\left[1 + \delta_L(t)\right].
\eq
A matter density perturbation impacts structure formation in two main ways: (i) it alters the total amount of mass that can participate in gravitational collapse and form bound structures and (ii) it modifies also the expansion rate of the spacetime inside the perturbation, which gives rise to a number of subtle points important to the setup of these separate universe simulations. We highlight some these points next and refer the reader to Refs.~\cite{li/hu/takada, 2014PhRvD..90j3530L, wagner/etal:2014, CFCpaper2, baldauf/etal:2015, response, lazeyras/etal, li/hu/takada:2016, 2018PhRvD..97l3526C, 2019MNRAS.488.2079B} for more details about them; in fact, some of the separate universe simulations that we use in this paper have been used already in Ref.~\cite{2019MNRAS.488.2079B} to study matter power spectrum response functions and calculate lensing bispectra and covariance matrices with baryonic effects taken into account. 

One of the effects of the modified expansion rate concerns the mapping between redshift and physical time, $z(t)$. The two cosmologies should be compared at the same physical time, but $N$-body codes like {\sc AREPO} specify the output epochs in terms of $z$, or scale factor $a = (1+z)^{-1}$. Consequently, outputs of the simulation of the fiducial cosmology at $z(t_{\rm out})$ should be compared with outputs of the separate universe simulation at $\tilde{z}(t_{\rm out}) \neq z(t_{\rm out})$. The starting redshift of the simulations should also be adjusted appropriately, although at such early times the amplitude of the perturbation is so small that the difference between $\tilde{z}(t_i)$ and $z(t_i)$ is numerically irrelevant.  Likewise, $N$-body codes typically take as input the parameters of the background cosmology evaluated at the time the scale factor is equal to unity. For the fiducial cosmology, this corresponds indeed to the present-day epoch, but this is not the case for the separate universe cosmologies. Thus, the background cosmological parameters that are used to run the separate universe simulations correspond to a time $\tilde{t}_0 \neq t_0$ defined by $\tilde{a}(\tilde{t}_0) = 1$. These are listed in Table \ref{table:params} for the case of $\delta_{L0} = +0.05$ (labeled ${\rm High}\ \delta_m$) and $\delta_{L0} = -0.05$ (labeled ${\rm Low}\ \delta_m$), where $\delta_{L0}$ is the present-day amplitude of the linear matter perturbation in the fiducial cosmology.\footnote{In the IllustrisTNG model, the time evolution of the amplitude of the UV radiation background is read from tabulated values as a function of redshift, not physical time $t$. These tabulated values should also be adjusted from $z(t)$ to $\tilde{z}(t)$ in the separate universe cosmologies. We did not perform this adjustment in our simulations, but we expect this to have only a negligible impact on the results we show here. This may not necessarily be the case, for instance, in studies of neutral/ionized gas fractions that may depend more sensitively on the detailed reionization history.}

{The modified expansion rate carries also implications to the halo finding criterion. Specifically, the FoF objects are found with a linking length $b$ times the mean distance between mass elements, but this distance is different in the comoving units of the fiducial and modified cosmologies. To ensure a consistent comparison between the structures found in the two cosmologies one must have $\tilde{a}(t) \tilde{b} = a(t) b$, which guarantees that the haloes correspond to the same physical scales. Noting that $\tilde{a}(t) = a(t)\left[1 + \delta_L(t)\right]^{-1/3}$, it follows that $\tilde{b} = b \left[1 + \delta_L(t)\right]^{1/3}$.  It is interesting to observe that this rescaling of the $b$ parameter naturally preserves halo mass definitions in terms of enclosed mass overdensities $\Delta$, $M_{\Delta} = (4\pi/3)\Delta {\rho}_mR_{\Delta}^3$, where $R_\Delta$ is the radius from the halo center enclosing a density $\Delta$ times the background matter density. First, given that $\tilde{\rho}_m(t) = \rho_m(t)\left[1+\delta_L(t)\right]$, the same halo in the modified cosmology therefore corresponds to an overdensity $\tilde{\Delta} = \Delta \left[1+\delta_L(t)\right]^{-1}$. Second, using percolation theory arguments, Ref.~\cite{2011ApJS..195....4M} showed that $b^3 \propto \Delta^{-1}$ provides a good description of the relation between $b$ and the mean density enclosed by the FoF boundary. Putting these two results together yields the rescaling of $b$ derived above. The parameters in the {\sc SUBFIND} code need no changing as they simply control the size of search radii for local density maxima, but more importantly, are given in terms of the number of particle neighbours and not in terms of a mean interparticle distance.}

The changes to the Hubble parameter $h$ in the different cosmologies imply also some care in the conversion of units that involve factors of $h$. For example, in {\sc AREPO}, the box size of the fiducial ($L_{\rm box}$) and separate universe ($\tilde{L}_{\rm box}$) simulations is quoted in units of ${\rm Mpc}/h$ and ${\rm Mpc}/\tilde{h}$, respectively; in our simulations, we set $\tilde{L}_{\rm box} = L_{\rm box}(\tilde{h}/h)$, which ensures that the comoving size of the boxes coincide at all times. The shape of the initial linear matter power spectrum is the same between the fiducial and separate universe cosmologies as $\Omega_{m0}h^2 = \tilde{\Omega}_{m0}\tilde{h}^2$, and hence, the same power spectrum file can be used to generate the initial conditions. The only modification is at the level of the conversion of wavenumber and spectra units from $h/{\rm Mpc}$ and ${\rm Mpc}^3/h^3$ to ${\tilde h}/{\rm Mpc}$ and ${\rm Mpc}^3/{\tilde h}^3$. Further, the structure finding algorithms return quantities like masses in units of $M_{\odot}/h$ and $M_{\odot}/\tilde{h}$ in the fiducial and separate universe cosmologies, respectively. In all our results, we convert the numerical value of all such quantities to the same units with the $h$ factor of the fiducial cosmology (specifically, if $M$ is the numerical value of the mass of a halo in the separate universe cosmology in units of $M_{\odot}/\tilde{h}$, then the value in $M_\odot/h$ units is $M(h/\tilde{h})$).

From its contribution to the galaxy bias expansion in Eq.~(\ref{eq:biasexp_2}), the linear LIMD galaxy bias parameter $b_1$ is formally defined as
\bq\label{eq:b1def}
b_1(z) = \frac{{\rm d} \ln n_g(z)}{{\rm d}\delta_L(z)} \bigg|_{\delta_{L}(z)=0},
\eq
which we evaluate by first-order finite-differencing the output of the ${\rm Fiducial}$, ${\rm High}\ \delta_m$ and ${\rm Low}\ \delta_m$ simulations. Effectively, we measure $b_1$ as
\bq\label{eq:b1measure}
b_1(z, M) = \frac{b_1^{\rm High}(z, M) + b_1^{\rm Low}(z, M)}{2},
\eq
with
\bq
\label{eq:b1measurehigh}b_1^{\rm High}(z, M) = \frac{1}{\delta^{\rm High}_L(z)}\Big[\frac{N_g^{{\rm High}\ \delta_m}(z,M)}{N_g^{\rm Fiducial}(z,M)} - 1\Big] + 1, \\
\label{eq:b1measurelow}b_1^{\rm Low}(z, M) = \frac{1}{\delta^{\rm Low}_L(z)}\Big[\frac{N_g^{{\rm Low}\ \delta_m}(z,M)}{N_g^{\rm Fiducial}(z,M)} - 1\Big] + 1,
\eq
and where the $N_g(z,M)$ denote, in the corresponding simulation, the number of haloes/subhaloes at redshift $z$ whose mass falls in some bin centered at $M$, and $\delta^{\rm High}_L(z) > 0$ and $\delta^{\rm Low}_L(z) < 0$.\footnote{In Eqs.~(\ref{eq:b1measurehigh}) and (\ref{eq:b1measurelow}), the $+1$ terms on the right-hand side account for the relation between the Lagrangian $b_1^L$ and Eulerian $b_1$ bias parameters. By construction, the comoving volumes of the fiducial and separate universe simulations agree, and hence, the derivative of $\ln N_g(z)$ naturally yields the Lagrangian bias, i.e., the bias defined w.r.t.~the initial density perturbations. In this paper, we work with the Eulerian bias that is defined in Eq.~(\ref{eq:biasexp_2}) w.r.t.~the matter fluctuations at later times. The volume elements in Eulerian space ${\rm d}^3\vx$ and Lagrangian space ${\rm d}^3\vq$ are related as ${\rm d}^3\vx = \left[1 + \delta_L(z)\right]^{-1}{\rm d}^3\vq$, i.e., the long-wavelength perturbation causes the Eulerian volume of the fiducial and separate universe simulations to be different: $V_{\rm Eul.}^{\rm Sep.Uni.} = V_{\rm Eul.}^{\rm Fiducial} / [1 + \delta_L]$. Plugging this relation into the finite-difference of number densities in Eulerian space,
\bq\label{eq:b1E}
b_1 = \frac{1}{\delta^{\rm Sep.Uni.}_L}\Big[\frac{N_g^{\rm Sep.Uni.}/V_{\rm Eul.}^{\rm Sep.Uni.}}{N_g^{\rm Fiducial}/V_{\rm Eul.}^{\rm Fiducial}} - 1\Big],
\eq
yields $b_1 = b_1^L + 1$.
}
Our choice of $|\delta_{L0}| = 0.05$ is motivated by having a sufficiently large perturbation to have a measurable impact in the simulations, while maintaining negligible higher-order corrections to the first-order finite-difference expression. The values of $b_1^{\rm High}(z, M)$ and $b_1^{\rm Low}(z, M)$ should be the same theoretically, but in practice, numerical noise and binning effects may drive some differences. The simulations we use in this work were run only once for a single realization of the initial conditions, which prevents us from quoting errors in a robust statistical sense. {As a compromise, we take $|b_1^{\rm High}(z, M)$ - $b_1^{\rm Low}(z, M)|/2$ as our estimate of the error of our measurements. The error bars can also be estimated by assuming that the galaxy counts are Poisson distributed in each bin, in which case the statistical error on $b_1$ is given by $\sigma_{b_1}^{\rm Poisson} = 1/\sqrt{2N_g^{\rm Fiducial}}/\delta_L(z)$. This is however an overestimate of the true statistical error because it does not take into account the correlation between $N_g^{\rm Fiducial}$ and $N_g^{\rm Sep.Uni.}$ that arises from the fact that the fiducial and separate universe simulations are run using the same phases of the initial conditions. Indeed, Ref.~\cite{baldauf/etal:2015} shows that the errors on $b_1$ estimated with Jackknife resampling using boxes with size $L = 1600\ {\rm Mpc}/h$ are instead well described by half of the Poisson expectation. We have explicitly checked that our error bars and $\sigma_{b_1}^{\rm Poisson}/2$ are comparable, which convinces us that our conclusions are not critically dependent on our estimate of measurement uncertainties.}

In Eqs.~(\ref{eq:b1measure})-(\ref{eq:b1measurelow}), we have explicitly considered objects binned in mass, but this can be straightforwardly generalized to any desired galaxy property. We shall do that below in Secs.~\ref{sec:biasgrc} and \ref{sec:biasbha} when we study the dependence of the galaxy bias on color and mass accretion rate of the galaxy black holes.

\subsection{The $\mathcal{A}_s$ Separate Universe}
\label{sec:sepuniAs}

The presence of PNG of the local type induces a non-vanishing bispectrum (Fourier transform of the three-point correlation function) in the primordial gravitational potential $\phi(\vx)$. This bispectrum peaks in the so-called squeezed-limit, which describes the coupling of long-wavelength perturbations with the power spectrum of two short-scale modes. In practice, this results in a modulation of the amplitude of the small-scale primordial scalar power spectrum by long-wavelength perturbations $\phi(\vx)$, which impacts the subsequent formation of structure that takes place inside such perturbations. Specifically, it can be shown that the primordial potential power spectrum evaluated locally around $\vx$ can be written as (see e.g.~Sec.~7.1.2 of the review Ref.~\cite{biasreview})
\bq\label{eq:Pks}
P_{\phi\phi}(k_{\tiny \rm short}, z | \vx) = P_{\phi\phi}(k_{\tiny \rm short}, z) \big[1 + 4\fnl\phi(\vx)\big],
\eq
where we have denoted the wavenumbers by $k_{\tiny \rm short}$ to emphasize that these are short-scale modes compared to the wavelength of the $\phi(\vx)$ perturbation. Similarly to the case of the total matter density perturbations then, galaxies forming inside a sufficiently long-wavelength perturbation regard it as a spatially uniform change to the variance of the small-scale fluctuations, i.e., they form as if in a separate cosmology with modified amplitude of the primordial power spectrum \cite{dalal/etal:2008, slosar/etal:2008}
\bq\label{eq:tildeAs}
\tilde{{\cal A}}_s = \mathcal{A}_s\left[1  + \delta \mathcal{A}_s\right], \ \ \ \ \ \ \ \ {\rm with} \ \ \ \ \ \ \ \delta \mathcal{A}_s = 4\fnl\phi_L
\eq
and where $\phi_L$ is the amplitude of the long-wavelength potential perturbation. 

Concerning the setup and analysis of the simulations, the only step that differs compared to the fiducial cosmology is at the level of the generation of the initial conditions, which should be generated using the same power spectrum file with the amplitude multiplied by $\left[1  + \delta \mathcal{A}_s\right]$ (cf.~Table \ref{table:params}). In our results, we consider cases with $\delta \mathcal{A}_s = \delta \mathcal{A}_s^{\rm High}= +0.05$ (labeled ${\rm High}\ \mathcal{A}_s$) and $\delta \mathcal{A}_s = \delta \mathcal{A}_s^{\rm Low} = -0.05$ (labeled ${\rm Low}\ \mathcal{A}_s$), and we evaluate $b_\phi$ as
\bq\label{eq:bphidef}
b_\phi(z,M) = 4 \frac{{\rm d} \ln n_g(z)}{{\rm d}\delta \mathcal{A}_s} \bigg|_{\delta \mathcal{A}_s=0};
\eq
the factor of $4$ arises because $b_\phi$ multiplies $\O(\vx) = \fnl\phi(\vx)$ in the galaxy bias expansion of Eq.~(\ref{eq:biasexp_2}), but the rescaling of the amplitude of the small scale power spectrum is $\delta \mathcal{A}_s = 4\fnl\phi_L$. Analogously to $b_1$, we evaluate $b_\phi$ as 
\bq\label{eq:bphimeasure}
b_\phi(z, M) = \frac{b_\phi^{\rm High}(z, M) + b_\phi^{\rm Low}(z, M)}{2},
\eq
where
\bq
b_\phi^{\rm High}(z, M) = \frac{4}{\delta \mathcal{A}_s^{\rm High}}\Big[\frac{N_g^{\rm High \mathcal{A}_s}(z,M)}{N_g^{\rm Fiducial}(z,M)} - 1\Big], \\
b_\phi^{\rm Low}(z, M) = \frac{4}{\delta \mathcal{A}_s^{\rm Low}}\Big[\frac{N_g^{\rm Low \mathcal{A}_s}(z,M)}{N_g^{\rm Fiducial}(z,M)} - 1\Big].
\eq
We estimate the error of our measurements of $b_\phi$ in the same way as $b_1$. In the calculation of $b_\phi$, $n_g$ and $N_g$ describe formally the abundance of galaxies in cosmologies with local PNG \cite{2000ApJ...541...10M, 2008JCAP...04..014L, 2011JCAP...08..003L}, but we evaluate them here using simulations with Gaussian distributed initial conditions. The impact of this is however unimportant in practice given the current observationally allowed values of $\fnl \lesssim 10$ \cite{2019arXiv190505697P}.

\section{Galaxy bias $b_1$ and $b_\phi$: numerical results}
\label{sec:biasresults}

In this section we present our measurements of the galaxy bias parameters $b_1$ and $b_\phi$ from the separate universe simulations of the IllustrisTNG model. We will show and discuss the results for both haloes and subhaloes, and as a function of total mass, total stellar mass, galaxy color and black hole mass accretion rate. In the section after this one, we analyse more carefully the distinct impact that matter perturbations and local-type PNG have on the SHMR, which will help explain the stellar mass results displayed in this section. 

\subsection{Dependence on total halo mass}
\label{sec:biasMh}

\begin{figure}[t!]
	\centering
	\includegraphics[width=\textwidth]{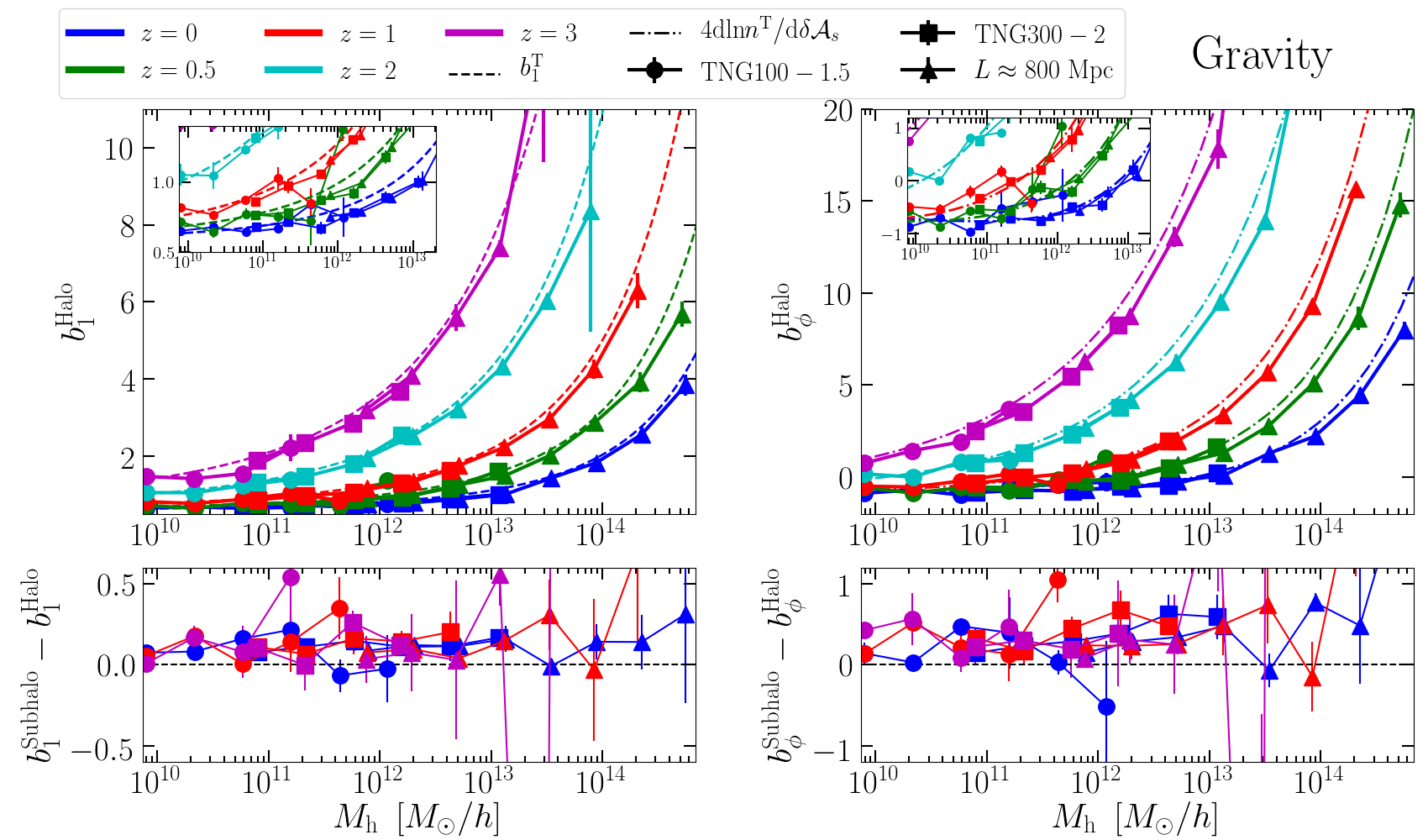}
	\caption{Linear (LIMD) bias $b_1$ (left) and local PNG bias $b_\phi$ (right) measured using the Gravity simulations as a function of total mass, $M_{\rm h}$. The symbols with errorbars connected with solid lines show the separate universe results from the TNG300-2 (squares), TNG100-1.5 (circles) and $L \approx 800\ {\rm Mpc}$ (triangles) resolutions, and the different colors indicate different redshifts, as labeled. The upper panels show the result for haloes and the lower panels shows the difference between the bias of subhaloes and haloes. In the upper left panel, the dashed curves show the prediction from the Tinker $b_1^{\rm T}$ fitting formula of Eq.~(\ref{eq:b1T}). In the upper right panel, the dot-dashed curves show the prediction of the separate universe argument applied to the Tinker halo mass function fitting formula (cf.~Eq.~(\ref{eq:bphidef_T})).}
\label{fig:b1bphi_dmo_totmass}
\end{figure}

Figure \ref{fig:b1bphi_dmo_totmass} shows the bias parameters $b_1$ (left) and $b_\phi$ (right) found in our Gravity simulations as a function of total mass $M_{\rm h}$ and at different redshifts, as labeled. In the upper left panel, we compare our $b_1$ measurements for haloes (symbols with errorbars) with the Tinker et al fitting formula of Ref.~\cite{2010ApJ...724..878T} shown by the dashed curves. The latter is given by
\bq\label{eq:b1T}
b_1^{\rm T}(z, M) = 1 - A\frac{\nu^a}{\nu^a + \delta_c^a} + B\nu^b + C\nu^c,
\eq
with $\nu = \delta_c/\sigma(z, M)$ and 
\bq\label{eq:sigmaM}
\sigma(z,M)^2 = \frac{1}{2\pi^2}\int {\rm d}k\,k^2 P^{\rm L}_{mm}(k,z) \tilde{W}^2(k, R(M)),
\eq
where $P^{\rm L}_{mm}$ is the linear matter power spectrum, $\tilde{W}(k, R(M)) = 3\left({\rm sin}(kR) - kR{\rm cos}(kR)\right)/\left(kR\right)^3$ and $R(M) = \left(3M/(4\pi\bar{\rho}_{m0})\right)^{1/3}$. {The fitting formula describes our $b_1$ measurements to within $\approx 5\%-10\%$ at all redshifts and mass scales shown.} This is in line with the expected accuracy of the formula for these redshifts (cf.~Fig.~1 of Ref.~\cite{2010ApJ...724..878T}); one should keep in mind as well that Ref.~\cite{2010ApJ...724..878T} calibrates the formula using haloes found with a spherical overdensity (SO) algorithm (with SO mass definitions; we adopt the $\Delta = 200$ values of Table 2 in Ref.~\cite{2010ApJ...724..878T}), whereas in this paper we deal with FoF haloes (with FoF mass definitions).

The upper right panel of Fig.~\ref{fig:b1bphi_dmo_totmass} compares the corresponding results for $b_\phi$ with a theoretical prediction obtained by applying the separate universe argument to fitting formulae of the halo mass function. Specifically, we adopt the Tinker et al. \cite{2008ApJ...688..709T} fitting formula for the differential halo mass function
\bq\label{eq:tinker_mf}
\frac{{\rm d}n^{\rm T}(M)}{{\rm d}M} &=& f^{\rm T}(\sigma) \frac{\bar{\rho}_{m0}}{M} \frac{{\rm d} \ln \sigma^{-1}}{{\rm d}M}, \\
f^{\rm T}(\sigma) &=& A\left[\left(\frac{\sigma}{b}\right)^{-a} + 1\right] {\rm exp}\left[-c/\sigma^2\right],
\eq
with $A = 0.186$, $a = 1.47$, $b=2.57$, $c = 1.19$ being the best-fitting parameters to the abundance of SO haloes in gravity-only simulations. The number of haloes $n^{\rm T}(M)$ within some mass bin centered at $M$ can be obtained by integrating Eq.~(\ref{eq:tinker_mf}), which depends on the amplitude of the linear matter power spectrum via $\sigma(z, M)$ (cf.~Eq.~(\ref{eq:sigmaM})). The theoretical prediction shown by the dot-dashed curve is then simply obtained using Eq.~(\ref{eq:bphidef}), but replacing the number counts measured from the simulations with those predicted by $n^{\rm T}(M)$:
\bq\label{eq:bphidef_T}
b^{\rm T}_\phi(z,M) = 4 \frac{{\rm d} \ln n^{\rm T}(M, z)}{{\rm d}\delta \mathcal{A}_s} \bigg|_{\delta \mathcal{A}_s=0}.
\eq
This equation is exact and it should be a good fit to the simulation results provided $n^{\rm T}$ captures well the impact of $\mathcal{A}_s$ on halo abundances {(recall the caveat that we deal with FoF haloes while $n^{\rm T}$ is a fitting function calibrated to the abundance of spherical-overdensity objects)}. Our results display however a trend for Eq.~(\ref{eq:bphidef_T}) to overpredict by about $10 - 20\%$ the measured $b_\phi$ above a certain halo mass; at $z=0$, $z=1$ and $z=2$, this mass scale is approximately $5\times 10^{13}\ M_{\odot}/h$,  $5\times 10^{12}\ M_{\odot}/h$ and $10^{11}\ M_{\odot}/h$, respectively.  

The lower panels of Fig.~\ref{fig:b1bphi_dmo_totmass} show the difference of the values of $b_1$ and $b_\phi$ between subhaloes and haloes, which reveal a trend for the subhaloes to be more biased than the haloes of the same mass. This is as expected because both $b_1$ and $b_\phi$ are growing functions of halo mass and the subhaloes reside inside more massive haloes. The noise in this measurement does not permit to discern any redshift- and mass-dependence of the difference, which is generically higher for $b_\phi$ ($b_\phi^{\rm Subhalo} - b_\phi^{\rm Halo} \approx 0.10-0.40$) than for $b_1$ ($b_1^{\rm Subhalo} - b_1^{\rm Halo} \approx 0.10-0.20$).

The upper panels of Fig.~\ref{fig:bphi_vs_b1_totmass_stemass} show the $b_\phi$ estimated from our simulations plotted against the corresponding $b_1$ measured in the same total mass bin for both haloes and subhaloes, and for different redshifts, as labeled (the lower panels correspond to stellar mass selection and are discussed in the next subsection). The universality relation $b_\phi = 2\delta_c(b_1-1)$ is shown by the solid black line, {which for $b_1 \gtrsim 1.5$ overpredicts slightly the measured $b_\phi$. This departure from the universality relation is well known from the literature \cite{grossi/etal:2009, desjacques/seljak/iliev:2009, 2010MNRAS.402..191P, 2010JCAP...07..013R, 2011PhRvD..84h3509H, scoccimarro/etal:2012, 2012JCAP...03..002W, baldauf/etal:2015, 2017MNRAS.468.3277B}. The grey band spans the area covered by a variant of the relation $b_\phi = q \times 2\delta_c(b_1-1)$ with $q \in \left[0.7, 0.9\right]$, which is in accordance with the results from past works. Note also that the size of the departure from the universality relation can depend on the halo mass definition \cite{2017MNRAS.468.3277B}.}

Although not shown, we have also explicitly verified that the total halo mass dependence of $b_1$ and $b_\phi$ measured from the Hydro simulations is perfectly consistent with those shown in Figs.~\ref{fig:b1bphi_dmo_totmass} from the TNG100-1.5 and TNG300-2 Gravity simulations. For $b_1$, this corroborates the previous findings from Ref.~\cite{2018MNRAS.475..676S} with IllustrisTNG of a negligible impact of baryonic effects on $b_1$ (estimated there using the large-scale limit of the ratio of the halo to matter power spectra). We find with this check that the total mass dependence of $b_\phi$ is also negligibly affected by baryonic effects.

\subsection{Dependence on stellar mass}
\label{sec:biasM*}

\begin{figure}[t!]
	\centering
	\includegraphics[width=\textwidth]{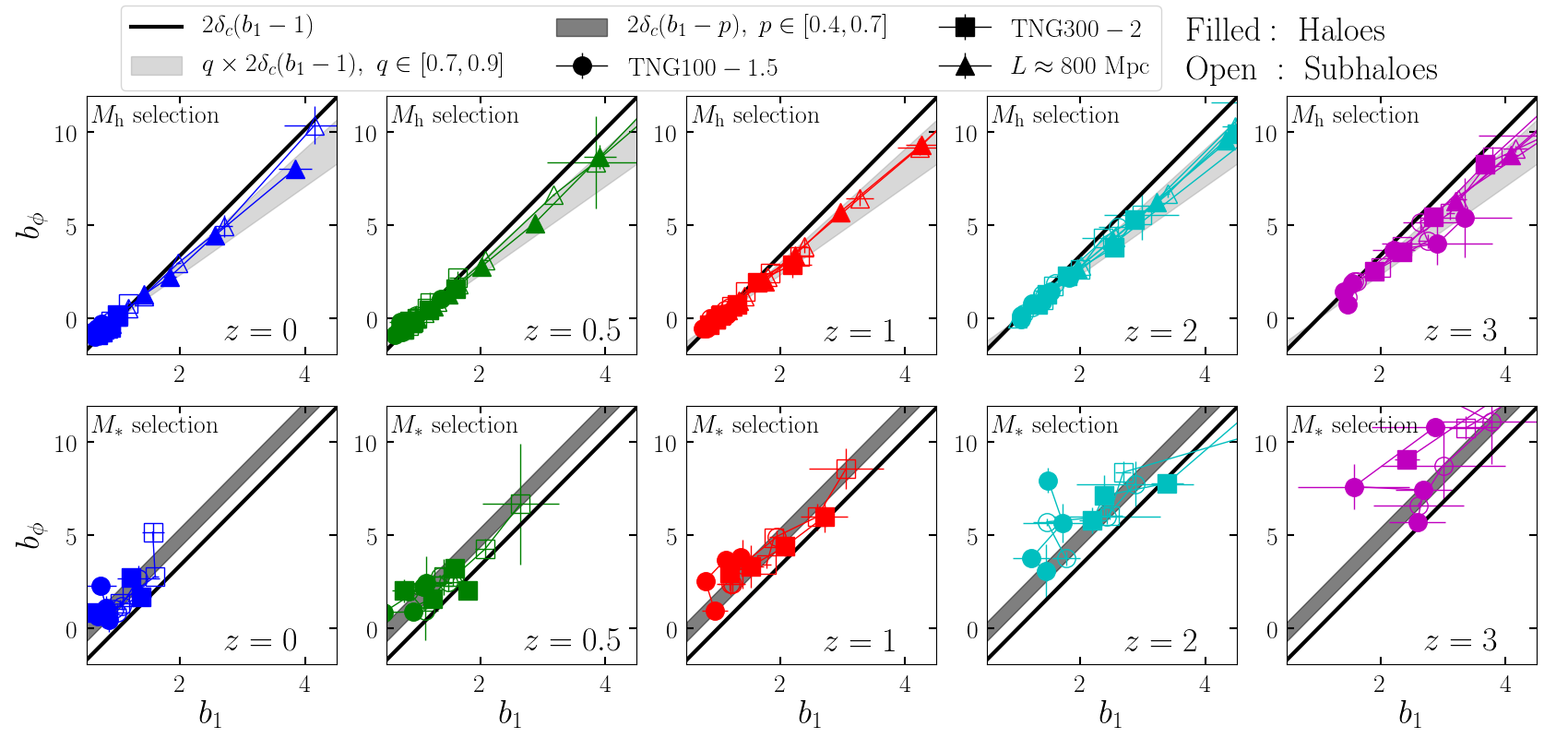}
	\caption{Local PNG bias $b_\phi$ plotted against the linear (LIMD) bias $b_1$ measured from the TNG100-1.5 (circles), TNG300-2 (squares) and $L \approx 800\ {\rm Mpc}$ (triangles) resolutions at different redshifts, as labeled. The upper and lower panels show the result for total halo mass and stellar mass selection, respectively; {each point shows $b_\phi$ and $b_1$ measured in the same mass bin (the total and stellar mass bins can be read from Figs.~\ref{fig:b1bphi_dmo_totmass} and \ref{fig:universality_b1bphi_stemass_groups}, respectively). The filled symbols are for haloes and the open symbols for subhaloes.} The solid black line marks the prediction from the universality relation $b_\phi = 2\delta_c(b_1-1)$,  {the light grey band in the upper panels marks the area covered by $b_\phi = q\times 2\delta_c(b_1-1),\ q \in \left[0.7, 0.9\right]$,} and the dark grey band in the lower panels shows the area covered by $b_\phi = 2\delta_c(b_1-p),\ p \in \left[0.4, 0.7\right]$.}
\label{fig:bphi_vs_b1_totmass_stemass}
\end{figure}

\begin{figure}[t!]
	\centering
	\includegraphics[width=\textwidth]{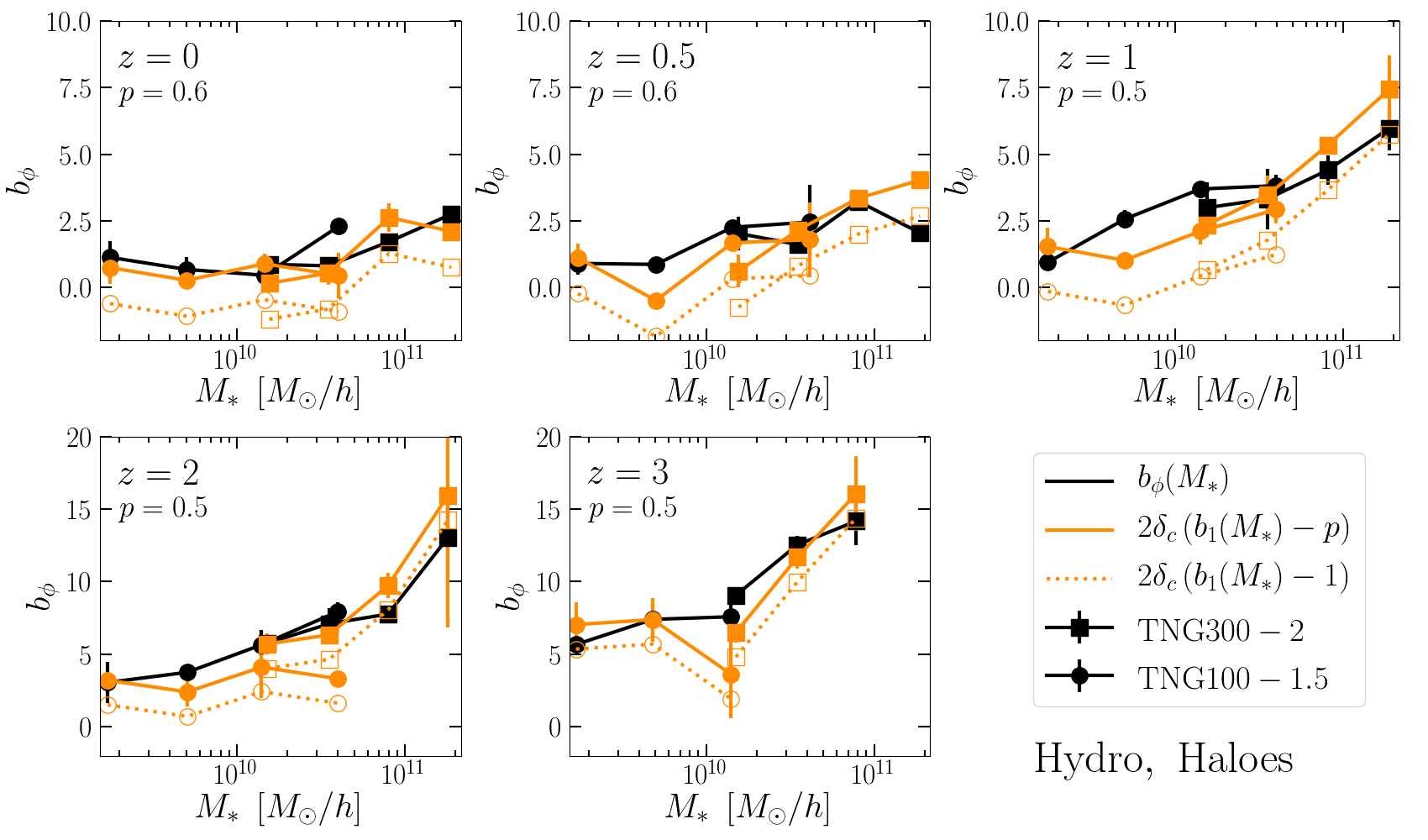}
	\caption{Local PNG bias $b_\phi$ measured for haloes (black symbols) as a function of their total stellar mass, $M_*$, and for different redshifts and resolutions, as labeled. The open orange symbols show the universality relation prediction using the corresponding $b_1(M_*)$ measurements. The filled orange symbols show the prediction of a modification of the universality relation, $b_\phi = 2\delta_c(b_1 - p)$; the value of $p$ is indicated in each panel.}
\label{fig:universality_b1bphi_stemass_groups}
\end{figure}

\begin{figure}[t!]
	\centering
	\includegraphics[width=\textwidth]{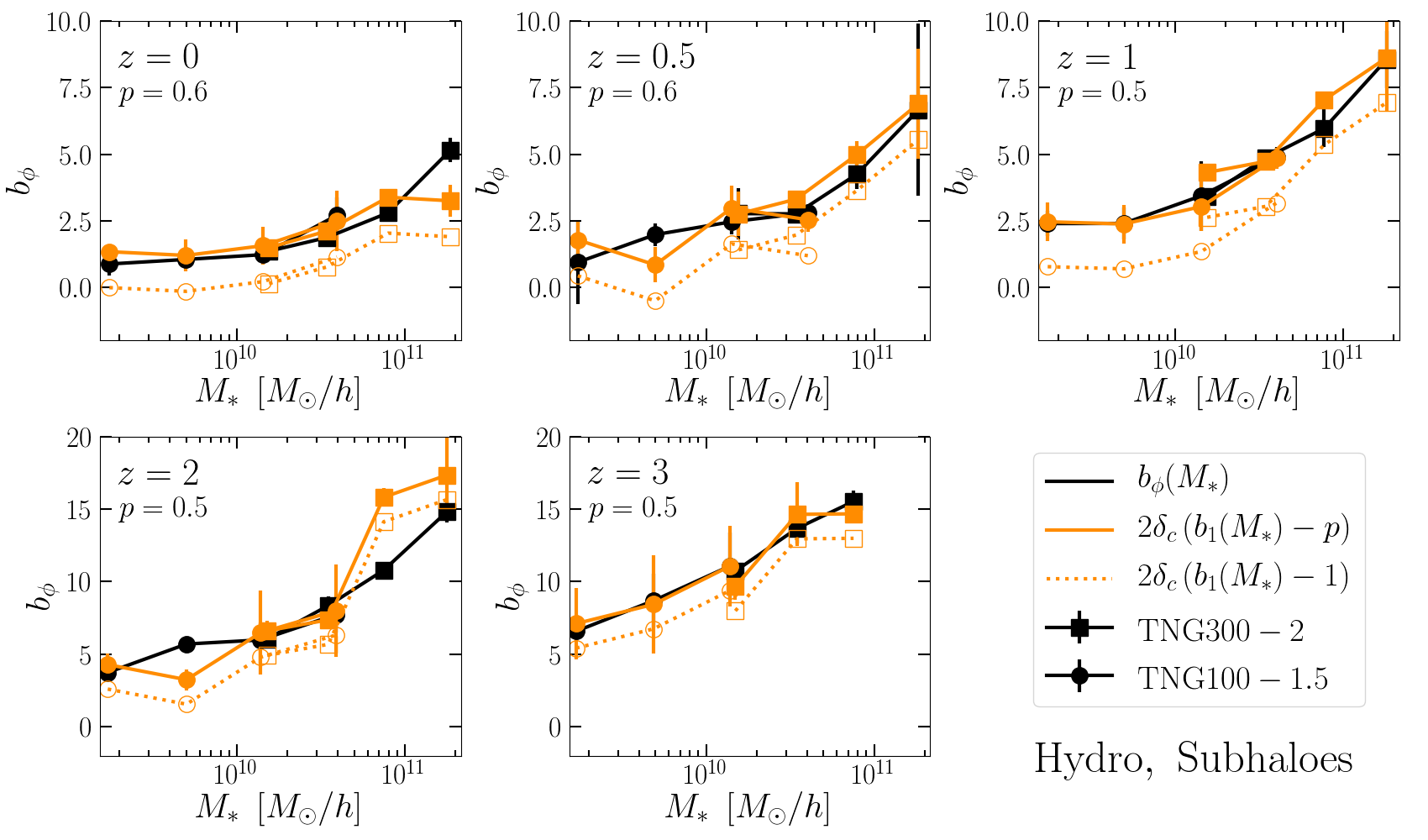}
	\caption{Same as Fig.~\ref{fig:universality_b1bphi_stemass_groups}, but for galaxies (i.e., subhaloes with stars) instead of haloes. }
\label{fig:universality_b1bphi_stemass_subhaloes}
\end{figure}

We now turn our attention to the bias of stellar-mass-selected haloes and galaxies and the performance of the universality relation $b_\phi = 2\delta_c(b_1-1)$ for such objects. This is addressed in the lower panels of Fig.~\ref{fig:bphi_vs_b1_totmass_stemass}, which show that, under stellar-mass selection, the universality relation underpredicts the $b_\phi(b_1)$ relation estimated from the simulations for all $b_1$ shown; i.e. stellar-mass selected objects show the opposite deviation from universality than total-mass selected haloes. 
In this case, we find that the variant $b_\phi = 2\delta_c(b_1-p)$ with $p \in \left[0.4, 0.7\right]$ provides a more accurate description of the simulation results. This form was put forward by Ref.~\cite{slosar/etal:2008} who argued that the relation with $p=1.6$, leading to a suppression relative to the universality relation, yields a better description of $b_\phi$ for objects whose host haloes had recently undergone a major merger; this was argued could be the case of most quasar host haloes, for example.\footnote{Using simulations with non-Gaussian initial conditions, Ref.~\cite{2010JCAP...07..013R} subsequently confirmed the dependence of $b_\phi$ on halo formation time at fixed halo mass, with older haloes being more biased than recently formed ones.} In our results, we find instead that $p \in [0.4, 0.7]$ works reasonably well in bringing the universality relation upwards into better agreement with the measured values of the $b_\phi(b_1)$ relation for both haloes and galaxies (recall, subhaloes with stars), as well as for all redshifts shown. The statistical precision of our measurements prevents us from carrying out a finer fitting of the parameter $p$ (which could be redshift dependent), but they nonetheless do robustly display a preference for $p<1$ for the stellar mass selected objects simulated with IllustrisTNG.  {The preference for $p \in [0.4, 0.7]$} is also robust to different stellar mass definitions: we checked this explicitly for stellar masses defined as the summed mass of (i) all stars in the haloes and galaxies (our default definition), (ii) all stars inside the stellar half-mass radius of galaxies and (iii) all stars inside twice the stellar half-mass radius of galaxies.

The breakdown of the universality relation for stellar-mass selected objects is displayed from another viewpoint in Figs.~\ref{fig:universality_b1bphi_stemass_groups} and \ref{fig:universality_b1bphi_stemass_subhaloes}, which show the stellar-mass dependence of $b_\phi$ for haloes and subhaloes (galaxies), respectively. The filled black symbols show the $b_\phi(M_*)$ measured from the simulations and the orange open symbols show the prediction of the universality relation using the corresponding $b_1(M_*)$ measurements (the interested reader can peek at Fig.~\ref{fig:modeling_b1_stemass_groups} below for the stellar mass dependence of $b_1$). As expected from Fig.~\ref{fig:bphi_vs_b1_totmass_stemass}, the universality relation underpredicts the measured $b_\phi(M_*)$ for both haloes and subhaloes, and at all redshifts and on all mass scales shown. To highlight the significance of this difference, note for example how in Fig.~\ref{fig:universality_b1bphi_stemass_subhaloes} the universality relation prediction approaches zero at $z<1$ for $M_* \lesssim 2\times 10^{10}\ M_{\odot}/h$, but the actual measured bias values remain quite sizeable $b_\phi \approx 1-3$. Adopting the universality relation for such galaxies in searches for local PNG therefore significantly underestimates the actual impact of $\fnl$ and leads to weaker constraints. The filled orange symbols in Figs.~\ref{fig:universality_b1bphi_stemass_groups} and \ref{fig:universality_b1bphi_stemass_subhaloes} show the prediction from the modified relation $b_\phi = 2\delta_c(b_1-p)$ with $p \in [0.5, 0.6]$, which indeed, describes the simulation results appreciably better.

\subsection{Dependence on galaxy color}
\label{sec:biasgrc}

\begin{figure}[t!]
	\centering
	\includegraphics[width=\textwidth]{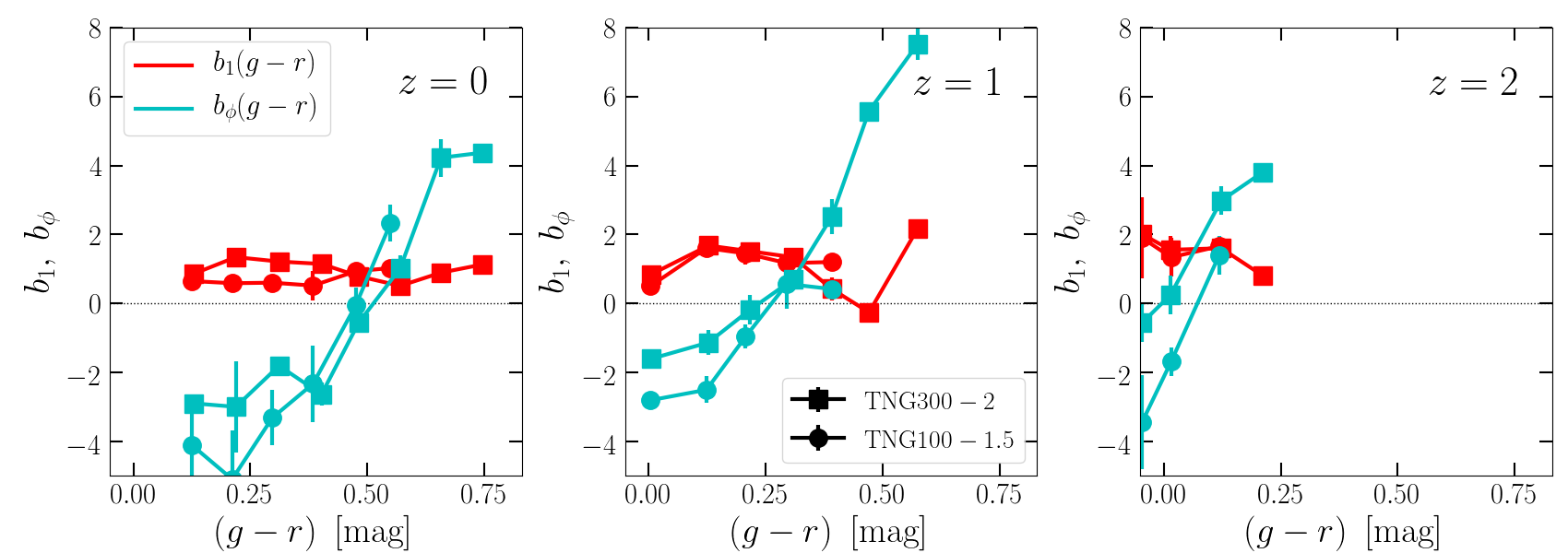}
	\caption{Galaxy bias parameters $b_1$ (red) and $b_\phi$ (cyan) as a function of $g-r$ color for galaxies; the different panels are for different redshifts, as labeled. Note that each bin in $g-r$ typically spans a wide range in mass.}
\label{fig:grcolor_peak}
\end{figure}

Our simulations with the IllustrisTNG model allow to study the dependence of the bias parameters $b_1$ and $b_\phi$ on galaxy color as well. Here, we analyse briefly the dependence on colors defined as the difference in dust-uncorrected $g$ and $r$ band luminosity due to all stars in the subhaloes; we do not account for the impact of dust attenuation on galaxy colors (see Ref.~\cite{Nelson:2017cxy} for an indepth discussion of galaxy colors in IllustrisTNG). Figure \ref{fig:grcolor_peak} shows the bias parameters $b_1$ and $b_\phi$ as a function of $g-r$ at $z=0$, $z=1$ and $z=2$. Their behavior is strikingly different, both qualitatively and quantitatively. At $z=0$ (left panel), $b_1$ stays approximately constant and equal to unity for all of the $g-r$ values shown. On the other hand, $b_\phi$ is always monotonically increasing with $g-r$: it is very negative at $g-r \approx 0.1$ ($b_\phi$ around $-3$ and $-4$), crosses zero at around $g-r = 0.5$ and becomes very positive $b_\phi = 4$ towards the {\it redder} end $g-r \approx 0.75$. At $z=1$ (middle panel), our results are compatible with a trend for $b_1$ to decrease from $b_1 \approx 2$ at $g-r \approx 0.1$ to $b_1 \approx 0$ at $g-r \approx 0.5$; at this redshift, $b_\phi$ displays a similar strong increase with $g-r$ to that at $z=0$ (the zero crossing happens at bluer values $g-r = 0.25$ and, at the redder end, it can grow up to $b_\phi = 8$). Finally, at $z=2$ (right panel), our results show that $b_1$ decreases and $b_\phi$ increases with $g-r$, but note that at this higher redshift the range in $g-r$ spanned by the galaxies is far smaller than at lower redshifts. 

The nontrivial dependence of $b_1$ and $b_\phi$ on galaxy color shown in Fig.~\ref{fig:grcolor_peak} motivates a more detailed analysis. {In Fig.~\ref{fig:grcolor_peak}, each $g-r$ bin covers a range in both total and stellar mass, which partly explains the different dependences of the bias parameters on mass and $g-r$. The response of the color-to-halo-mass relation is also an important factor in determining the shape of $b_1(g-r)$ and $b_\phi(g-r)$, similarly to as we discuss in the next section for the case of the stellar-to-halo-mass relation.} It would be further interesting to extend the analysis to luminosities in bands beyond $g$ and $r$, as well as including a treatment of attenuation by the dust distribution (which is itself potentially affected by overdensities and local PNG). We leave these and other developments to future work. 

\subsection{Dependence on black hole mass accretion rate}
\label{sec:biasbha}

\begin{figure}[t!]
	\centering
	\includegraphics[width=\textwidth]{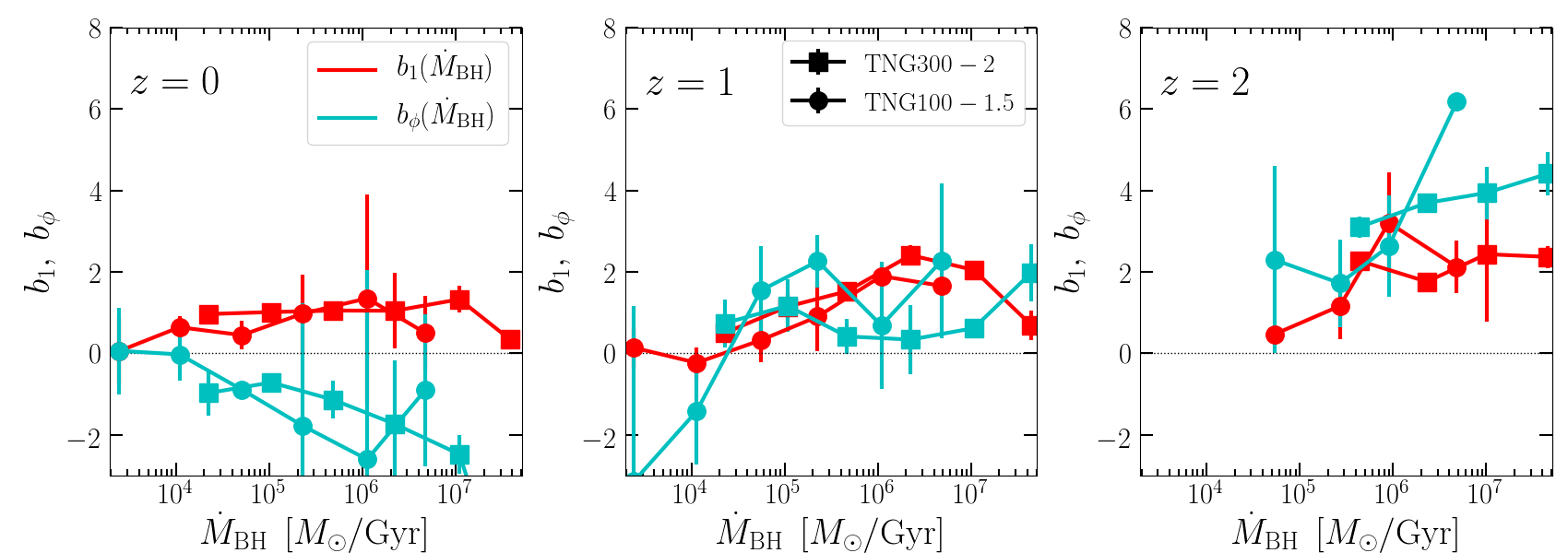}
	\caption{Same as Fig.~\ref{fig:grcolor_peak}, but for the instantaneous mass accretion rate of the galaxy's black holes, instead of $g-r$ color. Likewise, note that each bin in $\dot{M}_{\rm BH}$ typically spans a wide range in mass.}
\label{fig:bhmmdot_peak}
\end{figure}

The study of the dependence of galaxy bias on properties of their hosting black holes is particularly interesting to observational searches of local PNG, which usually make use of high redshift quasar samples to take advantage of the larger volumes they cover and probe sufficiently large scales. Figure \ref{fig:bhmmdot_peak} shows the dependence of $b_1$ and $b_\phi$ on the instantaneous mass accretion rate of the galaxy's black holes, $\dot{M}_{\rm BH}$. Like for the case of galaxy colors, the two bias parameters display rather different behaviors. At higher redshift $z=2$, $b_1$ and $b_\phi$ increase slightly with the black hole accretion rate, and at the high-$\dot{M}_{\rm BH}$ end, $b_\phi$ is larger than $b_1$: $b_\phi \approx 4$ vs.~$b_1 \approx 2$ for $\dot{M}_{\rm BH} \approx 10^7\ M_{\odot}/{\rm Gyr}$. With decreasing redshift, our results suggest a gradual and mild decrease in the value of $b_1$: by $z=0$, our results are consistent with $b_1 \approx 1$ for $\dot{M}_{\rm BH} \in \left[10^4, 10^7\right] M_{\odot}/{\rm Gyr}$. On the other hand, the value of $b_\phi$ decreases much more strongly with redshift: by $z=0$, it becomes negative with a trend for the galaxies with the faster accreting black holes to have more negative $b_\phi$.

Like for the dependence of the bias on galaxy color, we defer a more detailed investigation of the dependence of $b_1$ and $b_\phi$ on black hole properties to future work. We highlight once more the relevance of carrying out such a study, especially given the prominence of quasar power spectra in observational constraint studies of $\fnl$ \cite{slosar/etal:2008, 2011JCAP...08..033X, 2014PhRvD..89b3511G, 2014PhRvL.113v1301L, 2014MNRAS.441L..16G, 2015JCAP...05..040H, 2019JCAP...09..010C}.

\section{The impact of overdensities and local PNG on stellar and halo masses}
\label{sec:shmrimpact}

In this section, we study the response of the stellar-to-halo-mass-relation (SHMR) to the presence of long-wavelength perturbations of the matter field $\delta_m(\vx, z)$ and primordial gravitational potential perturbations with local PNG $\fnl\phi(\vx)$. We focus on haloes and begin by analysing the predictions of the IllustrisTNG model for the response functions themselves (our main conclusions here hold to the case of the galaxies as well). These are then later used as ingredients in modelling the stellar mass dependence of $b_1$ and $b_\phi$.

\subsection{Responses of the stellar-to-halo-mass relation}
\label{sec:responses}

\begin{figure}[t!]
	\centering
	\includegraphics[width=\textwidth]{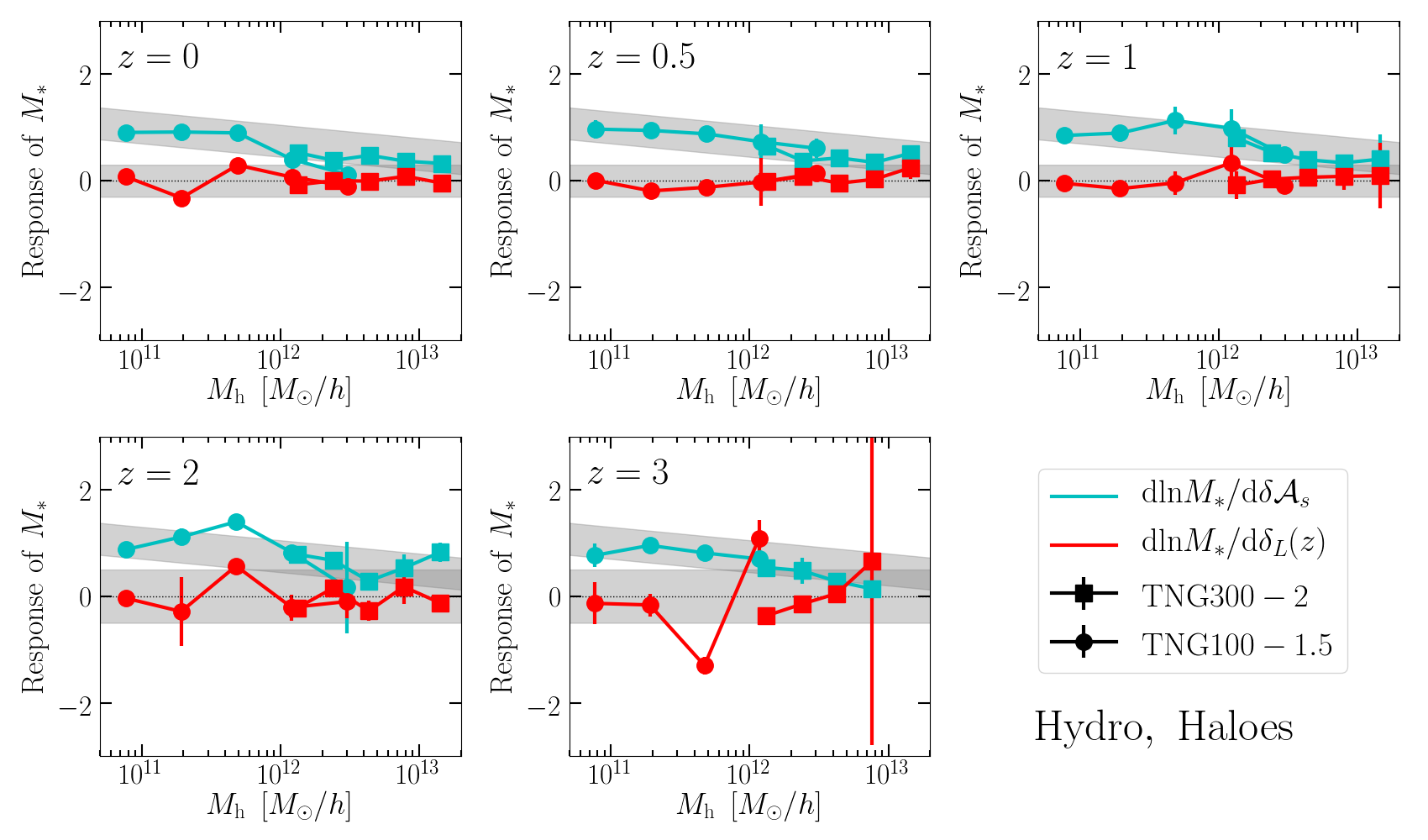}
	\caption{Response of the median stellar-to-halo-mass-relation (SHMR) to long-wavelength matter overdensities (red) and primordial potential perturbations with local PNG (cyan), at different redshifts, as labeled. The symbols show the response of the median stellar mass in fixed total halo mass bins. The grey bands indicate the regions covered by a simple model that we use in Sec.~\ref{sec:modeling} to describe the stellar mass dependence of $b_\phi$ and $b_1$. The corresponding response functions for galaxies (not shown) are effectively the same as those shown here for haloes.}
\label{fig:response_median_stemass_groups}
\end{figure}

\begin{figure}[t!]
	\centering
	\includegraphics[width=\textwidth]{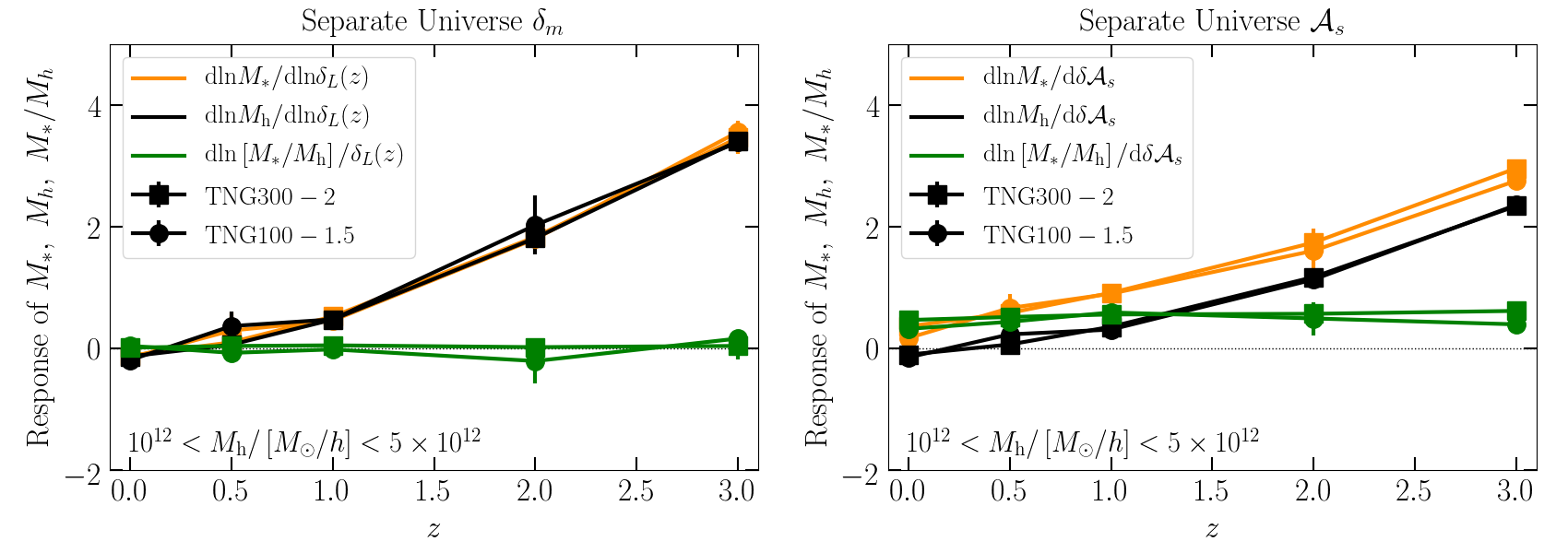}
	\caption{Responses of the total stellar mass (orange), total mass (black) and ratio of total stellar to total mass (green) of all haloes with mass $10^{12}\ M_{\odot}/h < M_{\rm h} < 5 \times 10^{12}\ M_{\odot}/h$. The result is shown as a function of redshift on the left for the response to matter overdensities and on the right for the response to local PNG.}
\label{fig:response_redshiftevo_Mstar_Mtot_MstaroMtot}
\end{figure}

\begin{figure}[t!]
	\centering
	\includegraphics[width=\textwidth]{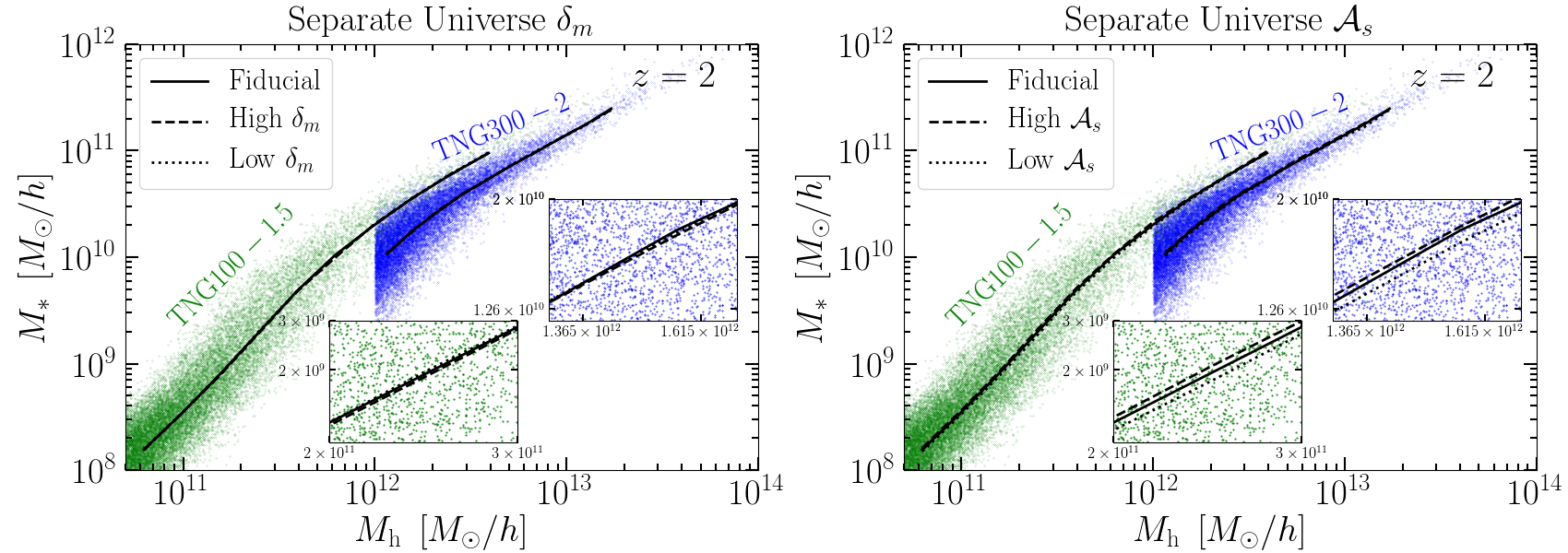}
	\caption{Stellar-to-halo-mass relation (SHMR) found in the TNG100-1.5 (green) and TNG300-2 (blue) simulations at $z=2$. The points show the relation for haloes in the fiducial cosmology. The black lines show the median stellar mass in bins of halo mass in the fiducial and separate universe cosmologies, as labeled. The left and right panels correspond, respectively, to the separate universes that mimic matter perturbations and primordial potential perturbations with local PNG. In each panel, the insets zoom into specific regions to actually visualize the different median relations. The relative difference between the black lines is effectively the $z=2$ response shown in Fig.~\ref{fig:response_median_stemass_groups}.}
\label{fig:scatter_Mstar_vs_Mtot}
\end{figure}

The response functions of the SHMR can be defined in analogy to the galaxy bias parameters by expanding the local SHMR in terms of these perturbations as
\bq\label{eq:SHMRexpansion}
M_*\left(M_{\rm h}, z | \vx\right) &=& M_*\left(M_{\rm h}, z \right) \left[1 + R_1^{M_*}(M_{\rm h}, z)\delta_m(\vx, z) + R_\phi^{M_*}(M_{\rm h}, z)\fnl\phi(\vx)\right],
\eq
where $M_*\left(M_{\rm h}, z | \vx\right)$ is the SHMR in a local volume around $\vx$ and $M_*\left(M_{\rm h}, z \right)$ is its cosmological average. The response coefficients are defined as
\bq\label{eq:SHMRresponses}
R_1^{M_*}(M_{\rm h}, z) &=& \frac{{\rm d} \ln M_*}{{\rm d}\delta_L(z)}\bigg|_{\delta_{L}(z)=0}, \\ 
R_\phi^{M_*}(M_{\rm h}, z) &=& 4 \frac{{\rm d} \ln M_*}{{\rm d}\delta \mathcal{A}_s} \bigg|_{\delta \mathcal{A}_s=0},
\eq
and can be evaluated via finite-differencing using our separate universe simulations. This is depicted in Fig.~\ref{fig:response_median_stemass_groups}, which shows the response of the median stellar mass in fixed total mass bins for the haloes found in the TNG100-1.5 and TNG300-2 simulations at different redshifts, as labeled. {The figure shows overall that matter perturbations and local PNG have a markedly different impact on the SHMR.} In the case of local PNG, a  boost of the amplitude of the primordial scalar power spectrum works to enhance the stellar mass that is found at fixed halo mass. This enhancement does not vary strongly with redshift (at least for $z<3$) and it is stronger at the lower-mass end of our measurements: ${{\rm d} \ln M_*}/{{\rm d}\delta \mathcal{A}_s}$ ranges from approximately unity to zero between $M_{\rm h} \sim 10^{11}\ M_{\odot}/h$ and $M_{\rm h} \sim 10^{13}\ M_{\odot}/h$. {In contrast, the response of the SHMR to a boost in matter density is effectively consistent with zero at all redshifts and mass scales shown. Both a boost in matter density and in amplitude of the primordial power spectrum work to enhance structure formation, and hence, the small size of the response of the SHMR to total matter perturbations may seem surprising as one could have expected this scenario to result also in enhanced star formation.} This seemingly unintuitive result can be traced back to the different impact that overdensities and local PNG have not only on stellar masses, but also on total halo masses, as we explain next. 

Figure \ref{fig:response_redshiftevo_Mstar_Mtot_MstaroMtot} shows the redshift evolution of the responses of the total stellar mass (orange), total halo mass (black) and corresponding ratios (green) for all  haloes found in a fixed halo mass bin $10^{12}\ M_{\odot}/h < M_{\rm h} < 5 \times 10^{12}\ M_{\odot}/h$. The responses to matter overdensities are depicted on the left and show that, at early times, overdense regions work to increase both the total mass and the stellar mass that is found in the chosen halo mass bin. This is due to two reasons: (i) there are more haloes in the mass bin and (ii) the individual haloes are themselves more massive. {Crucial to our discussion is the fact that the increment in total mass matches the increment in stellar mass, as shown by the equal amplitude of the black and orange lines, or by the small value of the response of the ratio of total stellar mass to total halo mass in green. In other words, long-wavelength total mass perturbations boost the rates of total mass accretion and star formation in the same way.} The right panel shows the same for the responses to local PNG, in which case the effects are likewise stronger at higher redshift, but importantly, the boost in stellar mass always exceeds the boost in total mass. 

To provide perhaps more intuition for the responses of the median SHMR depicted in Fig.~\ref{fig:response_median_stemass_groups}, we show in Fig.~\ref{fig:scatter_Mstar_vs_Mtot} the SHMR measured in our simulations at $z=2$. The points show the relation for haloes found in the TNG100-1.5 (green) and TNG300-2 (blue) simulations of the fiducial cosmology. The black curves show the median relation of the fiducial and corresponding separate universe cosmologies; the inset panels zoom into specific regions to visualize the actual changes. The lesson from Fig.~\ref{fig:response_redshiftevo_Mstar_Mtot_MstaroMtot} is that inside both positive long-wavelength matter perturbations $\delta_{L} > 0$ and positive primordial potential perturbations with local PNG ($\fnl\phi_L > 0$), the objects in the $M_* - M_{\rm h}$ plane of Fig.~\ref{fig:scatter_Mstar_vs_Mtot} are moved on average both rightwards (more total mass) and upwards (more stellar mass). Crucially, however, is the fact that in the case of matter overdensities, {the objects are moved rightwards and upwards by approximately the same amount. The net result is an effectively unchanged median stellar mass, when measured at fixed halo mass as indeed observed in Fig.~\ref{fig:response_median_stemass_groups}.} On the other hand, for the local PNG case, if $\delta \mathcal{A}_s > 0$, the objects are on average moved upwards more than they are moved rightwards in Fig.~\ref{fig:scatter_Mstar_vs_Mtot}, which explains the corresponding positive response of the SHMR shown in Fig.~\ref{fig:response_median_stemass_groups}.

We finish this subsection by highlighting the good agreement between the two resolution TNG100-1.5 and TNG300-2 predictions for the responses of the SHMR, which is remarkable given that predictions for their absolute values are still not converged at these numerical resolutions (cf.~Fig.~\ref{fig:scatter_Mstar_vs_Mtot}, but also Fig.~A2 of Ref.~\cite{Pillepich:2017jle} and Fig.~A1 of Ref.~\cite{Pillepich:2017fcc}). {This illustrates how the requirements on mass resolution (and consequently on computational resources) are less stringent in response studies, compared to studies of the corresponding absolute quantities.}

\subsection{Modeling $b_1$ and $b_\phi$ as a function of stellar mass}
\label{sec:modeling}

\begin{figure}[t!]
	\centering
	\includegraphics[width=\textwidth]{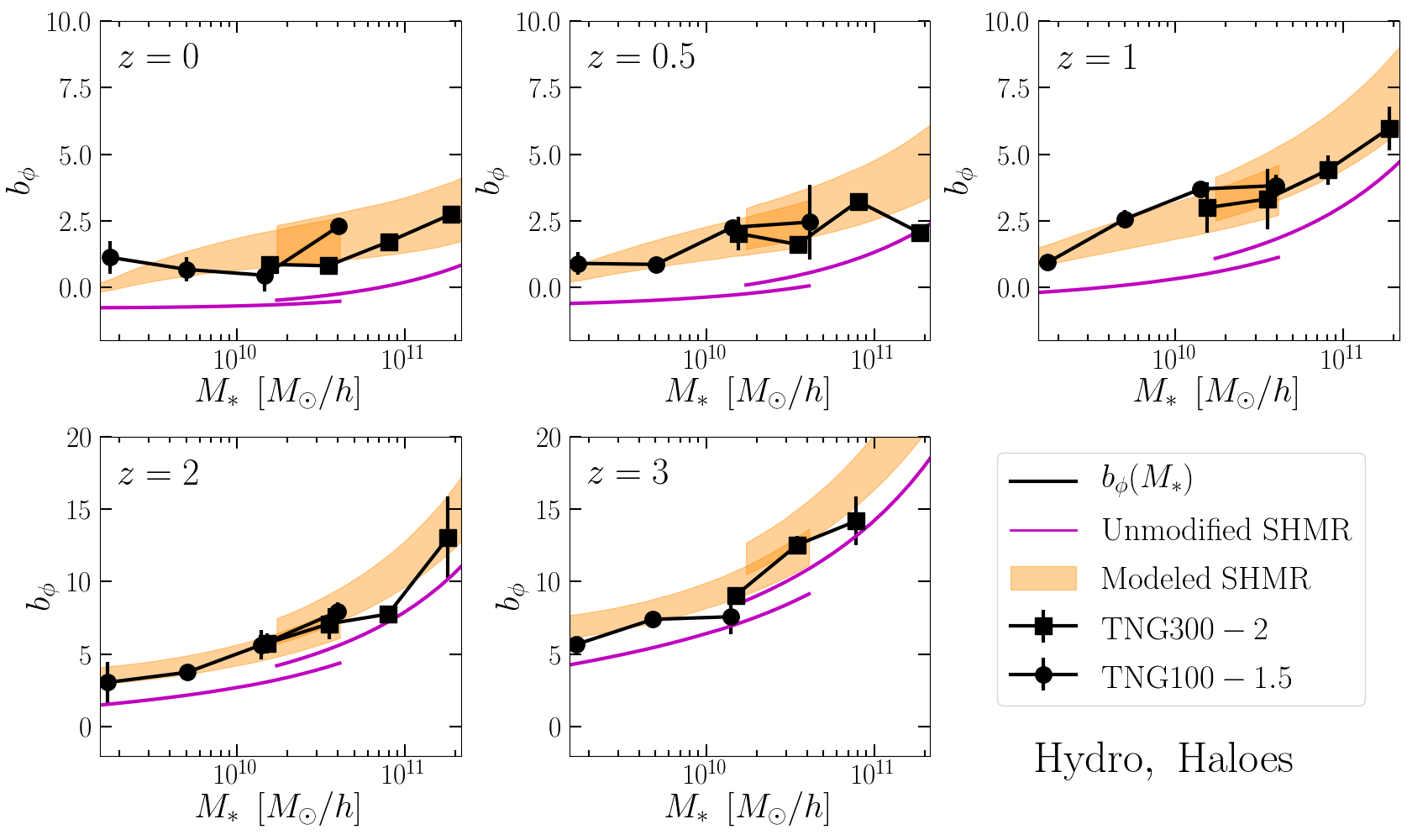}
	\caption{Local PNG bias $b_\phi$ measured for haloes (black symbols) as a function of their total stellar mass, $M_*$, and for different redshifts and resolutions, as labeled; these measurements are the same shown by the black symbols in Fig.~\ref{fig:universality_b1bphi_stemass_groups}. The orange bands show the predictions of our simple model using Eqs.~(\ref{eq:bphidef_T_model}), (\ref{eq:SHMRsepuniAs}) and (\ref{eq:SHMRmodel}) that takes into account the modifications to the SHMR by local PNG. The magenta curves show the outcome of the same model, but assuming an unmodified SHMR, i.e., $M_{\rm h}^{\rm Sep.Uni. \mathcal{A}_s}[M_*] \to M_{\rm h}^{\rm Fiducial}[M_*]$ in Eq.~(\ref{eq:bphidef_T_model}). In each panel, the two sets of model predictions correspond to using the $M_{\rm h}^{\rm Fiducial}[M_*]$ relation fitted to the TNG100-1.5 (left) and TNG300-2 (right) resolutions.}
\label{fig:modeling_bphi_stemass_groups}
\end{figure}

\begin{figure}[t!]
	\centering
	\includegraphics[width=\textwidth]{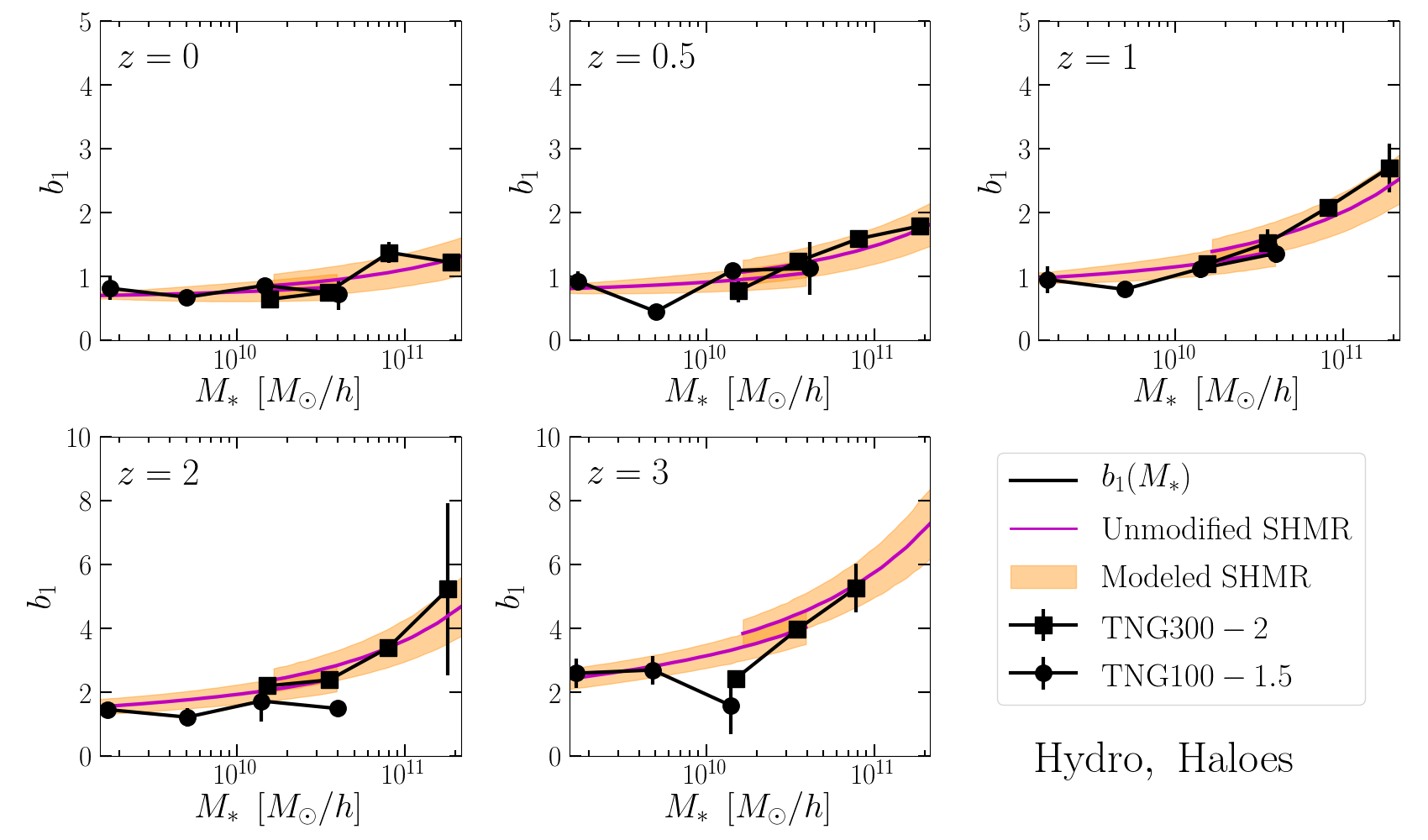}
	\caption{Same as Fig.~\ref{fig:modeling_bphi_stemass_groups}, but for $b_1$ instead of $b_\phi$.}
\label{fig:modeling_b1_stemass_groups}
\end{figure}

In this subsection, we attempt to model the stellar mass dependence of the bias parameters $b_1$ and $b_\phi$ by combining analytical formulae for the abundance of haloes with the responses of the SHMR discussed in the last subsection. We begin with $b_\phi$, which we model as
\bq\label{eq:bphidef_T_model}
b^{\rm T}_\phi(z, M_*) = \frac{1}{\delta \mathcal{A}_s}\left[\frac{n^{\rm T, Sep.Uni. \mathcal{A}_s}\big(M_{\rm h}^{\rm Sep.Uni. \mathcal{A}_s}[M_*]\big)}{n^{\rm T, Fiducial}\big(M_{\rm h}^{\rm Fiducial}[M_*]\big)} - 1\right],
\eq
where $n^{\rm T, Fiducial}$ and $n^{\rm T, Sep.Uni. \mathcal{A}_s}$ are the Tinker mass function formula evaluated for the fiducial and separate universe cosmologies, respectively; this equation corresponds simply to a modification of Eq.~(\ref{eq:bphidef_T}) to take $M_*$ as the argument via the halo-to-stellar-mass relation (the inverse of the SHMR) $M_{\rm h}(M_*)$, rather than $M_{\rm h}$ directly. For the fiducial cosmology, we describe $M_{\rm h}^{\rm Fiducial}[M_*]$ {by fitting a 3rd order polynomial to the median ${\rm log}_{10}M_h({\rm log}_{10}M_*)$ relation in our simulations}, whereas for the separate universe cosmology, we evaluate the same relation as 
\bq\label{eq:SHMRsepuniAs}
M_{\rm h}^{\rm Sep.Uni. \mathcal{A}_s}[M_*] = M_{\rm h}^{\rm Fiducial}[M_*(1 - R_\phi^{\rm M_*}\delta \mathcal{A}_s)],
\eq
i.e., at fixed stellar mass $M_*$, the associated halo mass in the separate universe cosmology is equal to the halo mass in the fiducial cosmology for $M_*$ rescaled by the SHMR response function $M_*(1 - R_\phi^{\rm M_*}\delta \mathcal{A}_s)$. We adopt a simple linear relation to model the response function 
\bq\label{eq:SHMRmodel}
R^{M_*}_\phi(M_h, z) = a \big[{\rm log}_{10}(M_h/M_0)\big] + c,
\eq
which brackets roughly our simulation measurements of $R_\phi^{M_*}$  for the parameters $M_0 = 10^{11}\ M_{\odot}/h$, $a = 0.25$ and $c = 0.7$ (lower bound) and $c=1.3$ (upper bound); this is the upper grey band in Fig.~\ref{fig:response_median_stemass_groups}.

The predictions from the model of Eqs.~(\ref{eq:bphidef_T_model}), (\ref{eq:SHMRsepuniAs}) and (\ref{eq:SHMRmodel}) are shown by the orange bands in Fig.~\ref{fig:modeling_bphi_stemass_groups}, which are able to describe very well the stellar-mass and redshift dependence of the $b_\phi$ measured from the simulations (black symbols). For comparison, the magenta lines show the outcome of the same model, but neglecting the impact of local PNG on the SHMR, i.e. $R_\phi^{M_*} \to 0$, or equivalently, $M_{\rm h}^{\rm Sep.Uni. \mathcal{A}_s}[M_*] \to M_{\rm h}^{\rm Fiducial}[M_*]$. Failing to incorporate the changes to the SHMR results in a significant underprediction of the value of $b_\phi$. This underprediction can be traced back to the breakdown of the universality relation discussed in Sec.~\ref{sec:biasM*}: the magenta curves in Fig.~\ref{fig:modeling_bphi_stemass_groups} are given by $b_\phi(M_{\rm h})$ where $M_{\rm h}$ is the halo mass corresponding to a given $M_*$, and as a result, they follow the universality prediction fairly closely. The universality relation $b_\phi = 2\delta_c(b_1 - 1)$ is therefore a poor description for stellar-mass selected objects to a large part because long-wavelength overdensities and primordial potential perturbations with local PNG impact the SHMR differently.

It is interesting to link our finding that $b_\phi(M_*)$ is enhanced by the positivity of the response $R_\phi^{M_*} > 0$ with the {\it assembly bias} finding of Ref.~\cite{2010JCAP...07..013R} that, at fixed mass, $b_\phi$ is higher in older than younger haloes. The boost in stellar mass induced by $\delta \mathcal{A}_s > 0$ could be at least partly a consequence of earlier halo formation times, or in other words, the behavior of $b_\phi$ for stellar-mass selected samples could be to some degree a manifestation of the secondary dependence of $b_\phi$ on the formation time of the host halo population. Indeed, Ref.~\cite{2010JCAP...07..013R} also finds that the galaxies in a stellar-mass selected sample constructed in Ref.~\cite{2007MNRAS.379.1143B} with semi-analytical modelling on the Millennium simulation do tend to populate older haloes and display, as a result, $b_\phi$ values that are higher than it would be inferred from the mass of their host haloes alone. A detailed look into these considerations is left for future work. 

Finally, concerning the stellar-mass dependence of $b_1$, we model it analogously to that of $b_\phi$ as
\bq\label{eq:b1def_T_model}
b^{\rm T}_1(z, M_*) = \frac{1}{\delta_L(z)}\left[\frac{n^{\rm T, Sep.Uni. \delta_m}\big(M_{\rm h}^{\rm Sep.Uni. \delta_m}[M_*]\big)}{n^{\rm T, Fiducial}\big(M_{\rm h}^{\rm Fiducial}[M_*]\big)} - 1\right],
\eq
where $n^{\rm T, Sep.Uni. \delta_m}$ represents a mass function prediction for the cosmologies that mimic long-wavelength total matter perturbations, i.e., cosmologies with spatial curvature. The Tinker fitting formula was not calibrated for those cosmologies, and so here we evaluate the corresponding abundance of haloes using the definition of the linear LIMD bias as
\bq\label{eq:nTsepunitrick}
n^{\rm T, Sep.Uni. \delta_m}(M_{\rm h},z) = n^{\rm T, Fiducial}(M_{\rm h},z)\Big[1 + b_1^{\rm T}(z, M_{\rm h})\delta_L(z)\Big].
\eq
The halo-to-stellar-mass relation in the separate universe cosmology is calculated as
\bq\label{eq:SHMRsepunidm}
M_{\rm h}^{\rm Sep.Uni. \delta_m}[M_*] = M_{\rm h}^{\rm Fiducial}[M_*(1 - R_1^{\rm M_*}\delta_L(z))],
\eq
with the response $R_1^{\rm M_*}$ modeled using the same relation as for $R_\phi^{\rm M_*}$ in Eq.~(\ref{eq:SHMRmodel}), {but centered around zero at all redshifts and mass scales. Specifically, we take $a=0$ for all redshifts (the value of $M_0$ is irrelevant in this case) and for $z = \{0, 0.5, 1, 2, 3\}$ we use, respectively, $c^{\rm upper} = \{0.3, 0.3, 0.3, 0.5, 0.5\}$ and $c^{\rm lower} = -c^{\rm upper}$, where the superscripts indicate the lower and upper bounds of the bottom grey bands in Fig.~\ref{fig:response_median_stemass_groups}.} The prediction of our modeling of $b_1(M_*)$ is shown by the orange bands in Fig.~\ref{fig:modeling_b1_stemass_groups}, which describe also well the $b_1(M_*)$ measured from the simulations. {Here, as expected, the SHMR response $R_1^{M_*}$ has negligible importance compared to the $b_\phi(M_*)$ case (cf.~magenta vs.~orange predictions in Figs.~\ref{fig:modeling_bphi_stemass_groups} and \ref{fig:modeling_b1_stemass_groups}) since $R_1^{\rm M_*}$ is much smaller than $R_\phi^{\rm M_*}$ (cf.~Fig.~\ref{fig:response_median_stemass_groups}).}

\section{Summary \& Conclusions}
\label{sec:summary}

The determination of the properties of the primordial density fluctuations is one of the main current goals in cosmology, with one of the key questions concerning the degree of primordial non-Gaussianity (PNG) of the distribution of these fluctuations. For the case of the so-called local-type PNG described by the amplitude parameter $\fnl$ (cf.~Eq.~(\ref{eq:fnl})), the current best constraints come from analysis of the CMB data $\fnl = -0.9 \pm 5.1\ (2\sigma)$, but upcoming surveys of the late-time galaxy distribution are expected to be able to probe $|\fnl| \lesssim 1$. The tightening of the bounds on $\fnl$ is of the utmost importance for fundamental physics as any non-zero detection of $\fnl$ would immediately rule out single-field models of inflation. A precise understanding of galaxy bias is however crucial to carrying out searches for $\fnl$ using the galaxy distribution because of their degenerate effects on statistics like the galaxy power spectrum (cf.~Eq.~(\ref{eq:Pgg})). Our main goal in this paper was precisely to study galaxy bias in the context of local PNG.

Specifically, in this paper, we have used cosmological hydrodynamical simulations with the IllustrisTNG galaxy formation model to study the linear LIMD galaxy bias $b_1$ and linear local PNG galaxy bias $b_\phi$. We have estimated these bias parameters using the separate universe technique, in which the local effect of the long-wavelength perturbations in a given fiducial cosmology is exactly mimicked by appropriate modifications to the cosmological parameters (cf.~Sec.~\ref{sec:sepuni}, Fig.~\ref{fig:idea} and Table \ref{table:params}). 

We have run separate universe simulations at two numerical resolutions, and both full hydrodynamical simulations with the IllustrisTNG galaxy formation physics model (dubbed Hydro), as well as gravity-only counterparts (dubbed Gravity). We have measured the bias parameters for both haloes (FoF groups) and subhaloes/galaxies ({\sc SUBFIND} substructures). We have focused our analysis on the dependence of $b_1$ and $b_\phi$ on total mass, stellar mass, galaxy color and black hole mass accretion rate. We have also studied the impact that matter overdensities and local PNG have on the stellar-to-halo-mass relation (SHMR). Our main results can be summarized as follows:

\begin{itemize}
\item For objects selected by their total mass, we recover the previously known result that the universality relation $b_\phi = 2\delta_c(b_1 - 1)$ overpredicts simulation results for $b_1 \gtrsim 1.5$ (cf.~Fig.~\ref{fig:bphi_vs_b1_totmass_stemass}).  We have also found no evidence of baryonic effects on $b_1$ and $b_\phi$ within the precision of our measurements. 

\item Our main result is that the universality relation underpredicts $b_\phi$ for stellar-mass selected objects (cf.~Figs.~\ref{fig:bphi_vs_b1_totmass_stemass}, \ref{fig:universality_b1bphi_stemass_groups} and \ref{fig:universality_b1bphi_stemass_subhaloes}). We find instead that $b_\phi(M_*) = 2\delta_c(b_1(M_*) - p)$ with $p \in [0.4, 0.7]$ provides a more accurate description.

\item The size and time evolution of $b_1$ and $b_\phi$ are very different for galaxies selected by color (cf.~Fig.~\ref{fig:grcolor_peak}) and black hole mass accretion rate (cf.~Fig.~\ref{fig:bhmmdot_peak}). This motivates work to establish the nontrivial relation between $b_1$ and $b_\phi$ for objects selected in terms of these two variables (or proxies thereof). 

\item {Long-wavelength matter overdensities boost the total mass and stellar mass of haloes equally.} In contrast, positive modulations of the primordial potential (with $\fnl > 0$) boost the mass in stars by more than they boost the total mass (cf.~Fig.~\ref{fig:response_redshiftevo_Mstar_Mtot_MstaroMtot}). These facts explain why the response of the median SHMR to local PNG is positive, {but the response to matter overdensities is effectively zero} (cf.~Fig.~\ref{fig:response_median_stemass_groups} and Sec.~\ref{sec:responses}).

\item By combining analytical formulae for the abundance of haloes as a function of total mass with simple modeling of the responses of the SHMR, we were able to reproduce the stellar-mass dependence of $b_\phi$ and $b_1$ (cf.~Figs.~\ref{fig:modeling_bphi_stemass_groups} and \ref{fig:modeling_b1_stemass_groups}). The incorporation of the modifications of the SHMR in this modeling is crucial to reproduce the measured $b_\phi$, and neatly illustrates that the poor performance of the universality relation under stellar-mass selection is at least partly due to the different response of the SHMR to overdensities and local PNG.
\end{itemize}

{Taken at face value, our finding that $b_\phi(M_*) = 2\delta_c(b_1(M_*) - p)$ with $p \in[0.4, 0.7]$ (instead of $p=1$) is a better description for stellar mass selected galaxy samples may represent good news for local PNG searches using galaxies. Concretely, $p=1$ is typically used in forecast studies of the constraining power on $\fnl$, but our results indicate that this underpredicts the amplitude of the true effect that local PNG would have on the galaxy power spectrum of IllustrisTNG galaxies (see e.g.~Ref.~\cite{2017PhRvD..95l3513D} for a detailed forecast study on $\fnl$ using stellar mass selected samples). The magnitude of this underprediction can be quite substantial depending on mass and redshift. For example, taking the case of our results for $M_* \approx 10^{10}\ M_{\odot}/h$ galaxies at $z=1$ in Fig.~\ref{fig:universality_b1bphi_stemass_subhaloes}, the universality relation predicts that the clustering of these objects is fairly insensitive to $\fnl$ (i.e., $|b_\phi| \lesssim 1$), but our results show that the measured bias is in fact about 3 times larger, $b_\phi \approx 3$. Adopting $b_\phi(M_*) = 2\delta_c(b_1(M_*) - p)$ in constraint/forecast analyses with $p$ marginalized over a range with $p < 1$ could therefore reveal that the prospects to detect non-zero $\fnl$ using galaxies may be more promising than previously thought.}

{This example highlights the critical role of the relation between $b_1$ and $b_\phi$ in $\fnl$ constraints. Recently, Ref.~\cite{2020arXiv200906622B} studied this in more detail and found that the bounds on $\fnl$ can indeed depend sensitively on the prior adopted for the parameter $p$. This provides strong motivation to extend our analysis beyond the IllustrisTNG model onto other state-of-the-art simulations of galaxy formation, and use the mean and scatter of the various predictions to inform priors on $b_\phi$ in constraint analyses/forecasts of $\fnl$. We note additionally that searches for local PNG using galaxies will rely also on samples selected by properties other than stellar mass, for which $b_\phi = 2\delta_c(b_1 - p)$ may not even be an adequate functional form (Figs.~\ref{fig:grcolor_peak} and \ref{fig:bhmmdot_peak} suggest this is the case for objects selected by color and black hole accretion rate). For these cases, future work should focus also on developing appropriate parametrizations of the $b_\phi(b_1)$ relation and/or identifying other ways to place priors on $b_\phi$.}

Our results demonstrate further that the relation between two or more galaxy bias parameters can encode interesting information on galaxy formation, namely, its coupling to the long-wavelength environment. One can hence entertain the idea to use sufficiently precise estimates of galaxy bias parameters from data to constrain galaxy formation. It would thus be interesting to extend our analysis to other galaxy bias parameters and to the response of other galaxy properties.


\acknowledgments
The simulations used in this work were run on the Cobra supercomputer at the Max Planck Computing and Data Facility (MPCDF) in Garching near Munich. AB, GC and FS acknowledge support from the Starting Grant (ERC-2015-STG 678652) ``GrInflaGal'' from the European Research Council.

\bibliographystyle{utphys}
\bibliography{REFS}

\end{document}